\documentclass[twocolumn]{aastex61}

\usepackage[percent]{overpic}

\usepackage{xspace}
\usepackage{moresize}

\newcommand{\LAT}{\emph{Fermi}-LAT\xspace}
\newcommand{\HESS}{H.E.S.S.\xspace}
\newcommand{\extragal}{extragalactic\xspace}

\newcommand{\sts}{$\sqrt{\mathrm{TS}}$\xspace}

\newcommand{\degree}{$^\circ$\xspace}
\newcommand{\ergps}{$\, \textrm{erg} \, \textrm{s}^{-1}$\xspace}
\newcommand{\ifu}{$\, \textrm{cm}^{-2} \, \textrm{s}^{-1}$\xspace}
\newcommand{\dfu}{$\, \textrm{TeV}^{-1} \, \textrm{cm}^{-2} \, \textrm{s}^{-1}$\xspace}

\begin{document}

\title{The 2HWC HAWC Observatory Gamma Ray Catalog}

\correspondingauthor{C. Rivi\`ere}
\email{riviere@umdgrb.umd.edu}

\AuthorCollaborationLimit=300
\author{A.U.~Abeysekara}
\affil{Department of Physics and Astronomy, University of Utah, Salt Lake City, UT, USA}

\author{A.~Albert}
\affil{Physics Division, Los Alamos National Laboratory, Los Alamos, NM, USA}

\author{R.~Alfaro}
\affil{Instituto de F\'{i}sica, Universidad Nacional Aut\'{o}noma de M\'{e}xico, Mexico City, Mexico}

\author{C.~Alvarez}
\affil{Universidad Aut\'{o}noma de Chiapas, Tuxtla Guti\'{e}rrez, Chiapas, Mexico}

\author{J.D.~\'{A}lvarez}
\affil{Universidad Michoacana de San Nicol\'{a}s de Hidalgo, Morelia, Mexico}

\author{R.~Arceo}
\affil{Universidad Aut\'{o}noma de Chiapas, Tuxtla Guti\'{e}rrez, Chiapas, Mexico}

\author{J.C.~Arteaga-Vel\'{a}zquez}
\affil{Universidad Michoacana de San Nicol\'{a}s de Hidalgo, Morelia, Mexico}

\author{H.A.~Ayala Solares}
\affil{Department of Physics, Michigan Technological University, Houghton, MI, USA}

\author{A.S.~Barber}
\affil{Department of Physics and Astronomy, University of Utah, Salt Lake City, UT, USA}

\author{N.~Bautista-Elivar}
\affil{Universidad Politecnica de Pachuca, Pachuca, Hidalgo, Mexico}

% ...

\begin{abstract}
  We present the first catalog of TeV gamma-ray sources realized with the
recently completed High Altitude Water Cherenkov Observatory (HAWC).
It is the most sensitive
wide field-of-view TeV telescope currently in operation, with a 1-year survey
sensitivity of $\sim$5--10\% of the flux of the Crab Nebula.
With an instantaneous field of view $>$1.5\,sr and $>$90\% duty cycle, it
continuously surveys and monitors the sky for gamma ray energies between hundreds GeV and tens of TeV.

HAWC is located in Mexico at a latitude of 19\degree North and was completed
in March 2015.
Here, we present the 2HWC catalog, which is the result of the first source
search realized with the complete HAWC detector.
Realized with 507\,days of data and
represents the most sensitive TeV survey to date for such a large fraction of the sky.
A total of 39 sources were detected, with an expected contamination of 0.5 due to 
background fluctuation.
Out of these sources, 16 are more than one degree away from any previously
reported TeV source.
The source list, including the position measurement, spectrum measurement, and
uncertainties, is reported.
Seven of the detected sources may be associated with pulsar wind nebulae, two
with supernova remnants, two with blazars, and the remaining 23 have no firm
identification yet.

\end{abstract}

\section{Introduction}
  \label{sec:intro}
  The High Altitude Water Cherenkov Observatory (HAWC) is a newly completed 
very high energy (VHE; $>$100\,GeV) gamma-ray observatory with a 1-year survey sensitivity
of $\sim$5--10\% of the flux of the Crab Nebula. The variation in sensitivity depends on 
the declination of the source under consideration over the observable sky, with
declinations between $-20$\degree and 60\degree for the present study.
Unlike imaging atmospheric Cherenkov telescopes (IACTs), such 
as H.E.S.S. \citep{2004APh....22..109A}, MAGIC \citep{2016APh....72...61A}, VERITAS \citep{2006APh....25..391H},
and FACT \citep{2011NIMPA.639...58A} which observe the Cherenkov light emitted
by the extensive air showers as they develop in the atmosphere,
HAWC detects particles of these air showers that
reach ground level, allowing it to operate continuously and observe 
an instantaneous field of view of $>$1.5\,sr.
Prior to this work, unbiased VHE surveys were conducted by the Milagro \citep{directintegration, 2004ApJ...608..680A} and ARGO \citep{2002APh....17..151A}
collaborations.
Compared to these previous surface arrays,
the sensitivity of HAWC is improved by more than an order of magnitude thanks to
a combination of large size, high elevation, and unique
background rejection capability. These features make HAWC ideally suited as a VHE survey instrument. 
High-sensitivity 
surveys of portions of the Galactic Plane have also been published by H.E.S.S. \citep{2006ApJ...636..777A}, MAGIC \citep{2006ApJ...638L.101A} and 
VERITAS \citep{2015arXiv150806684P}. At lower energies, the Large Area Telescope on the space-based 
Fermi Observatory (\LAT) has detected many thousands of Galactic and \extragal gamma-ray 
sources \citep{2015ApJS..218...23A}, but its small size limits its reach into the VHE band.
The work presented here is the most sensitive comprehensive sky survey carried out above 1\,TeV.

There are about 200 known VHE gamma-ray sources detected at high significance by a 
number of observatories  \citep[e.g. TeVCat catalog;][]{2008ICRC....3.1341W}.

Within the Galaxy, the VHE sources include pulsar wind nebula (PWNe), supernova
remnants (SNRs), binary systems, and diffuse emission from the Galactic plane.
The SNRs and PWNe represent the majority of the identified sources.
Most Galactic gamma-ray sources have power-law spectra consistent with 
shock acceleration of electrons, though there is considerable evidence for
gamma-ray production by hadronic cosmic rays interacting with matter.
Most Galactic sources are observed as spatially extended by IACTs \citep{2013arXiv1307.4690C}.

Beyond our galaxy, almost all known TeV sources are Active 
Galactic Nuclei (AGNs) and most of them are 
categorized as blazars. 
The TeV gamma-ray emission is generally observed to be 
variable and thought to originate from one or multiple regions of 
particle acceleration in the jet. 
While gamma-ray emission has been observed up to energies 
of about 10\,TeV for some blazars
\citep{2011ApJ...738...25A,2001A&A...366...62A}, the flux at and 
beyond such energies is strongly attenuated as a function of distance 
due to photon-photon interaction with the \extragal background light (EBL). Since 
the sensitivity of HAWC peaks around 10\,TeV (depending on the source spectrum
and declination, see Section~\ref{sec:performance} for details), where absorption of 
TeV photons through the infrared component of the EBL becomes severe,
the sensitivity of the HAWC survey to distant AGNs is relatively poor.

Many VHE sources are not unambiguously associated with 
objects identified at other wavelengths (a fifth of TeVCat sources are reported as unidentified). Further spectral and morphological studies are 
required to understand their origins and emission mechanisms.

In addition to a peak sensitivity at higher energies, the angular resolution
of HAWC is larger than the IACT's. Consequently,
comparison of source significance and flux with IACT observations requires careful examination. For example,
the HAWC instrument is relatively more sensitive to sources with harder energy spectra than softer ones, and to extended sources than pointlike sources. 
On the other hand, the surface detection method 
employed by HAWC permits continuous observation of the entire overhead sky,
both during the day and night and under all weather 
conditions.
For sources that transit through its field of view, HAWC typically accumulates
1500--2000\,hours/yr of total exposure.
Thus, above 10\,TeV where photon statistics are poor, HAWC achieves better sensitivity than even long-duration
observations by IACTs.

This paper presents a catalog of TeV gamma-ray sources resulting from a search
for significantly enhanced point and extended emission detected in the gamma-ray
sky maps of 17 months of HAWC data.
More detailed morphology studies will be the subject of future papers.

In Section \ref{sec:detector}, we describe the HAWC detector. Section \ref{sec:method} describes the analysis
of gamma-ray events and the construction of our source catalog.
Results and discussion are provided in Sections \ref{sec:results}, \ref{sec:discussion}, and \ref{sec:population}, and conclusions
and outlook in Section \ref{sec:outlook}.

\section{HAWC Detector}
  \label{sec:detector}
\begin{figure}
  \centering
  \begin{overpic}[trim={1.5cm 1cm 0.5cm 2cm},clip,width=\columnwidth]{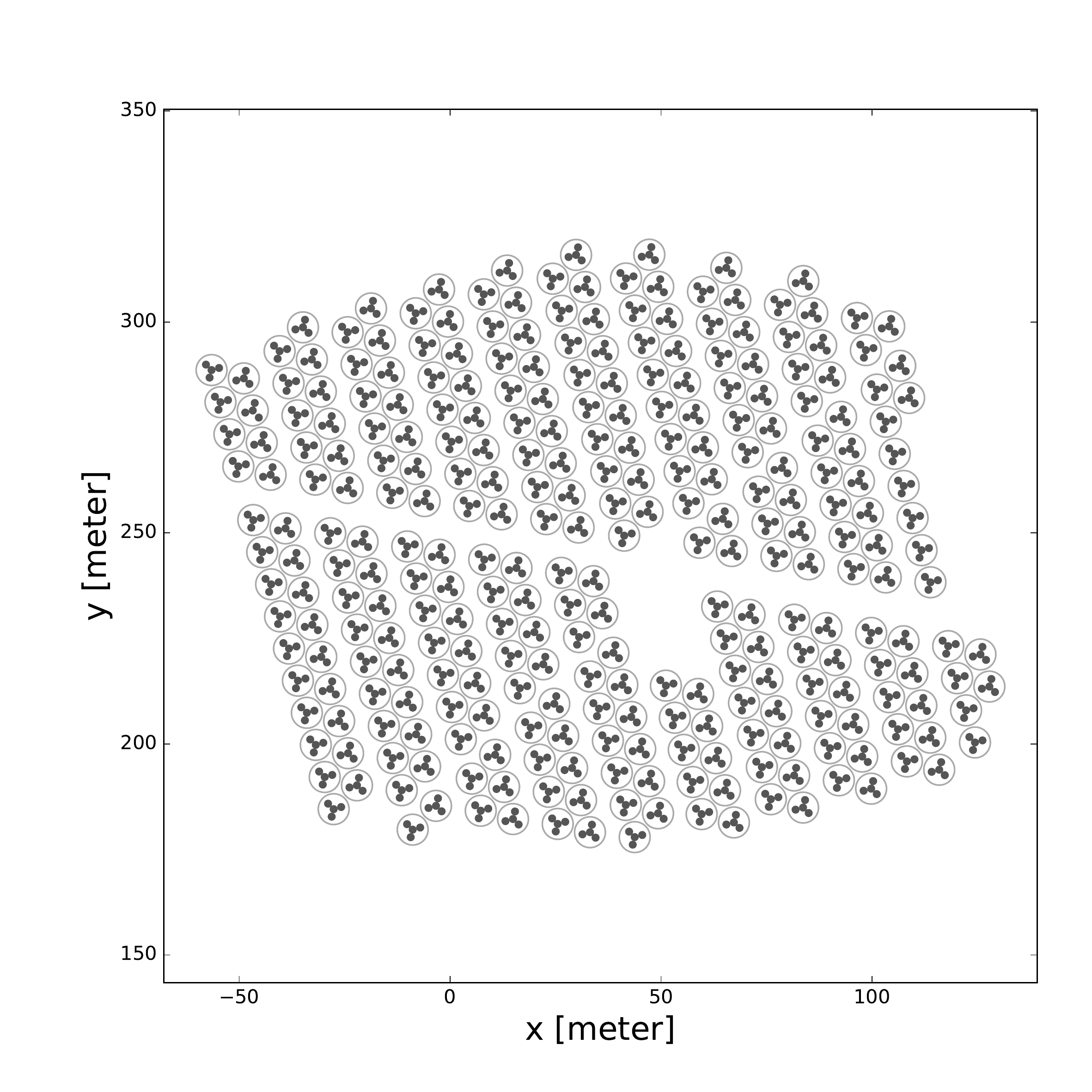}
     \put(81,77){\includegraphics[width=0.15\columnwidth]{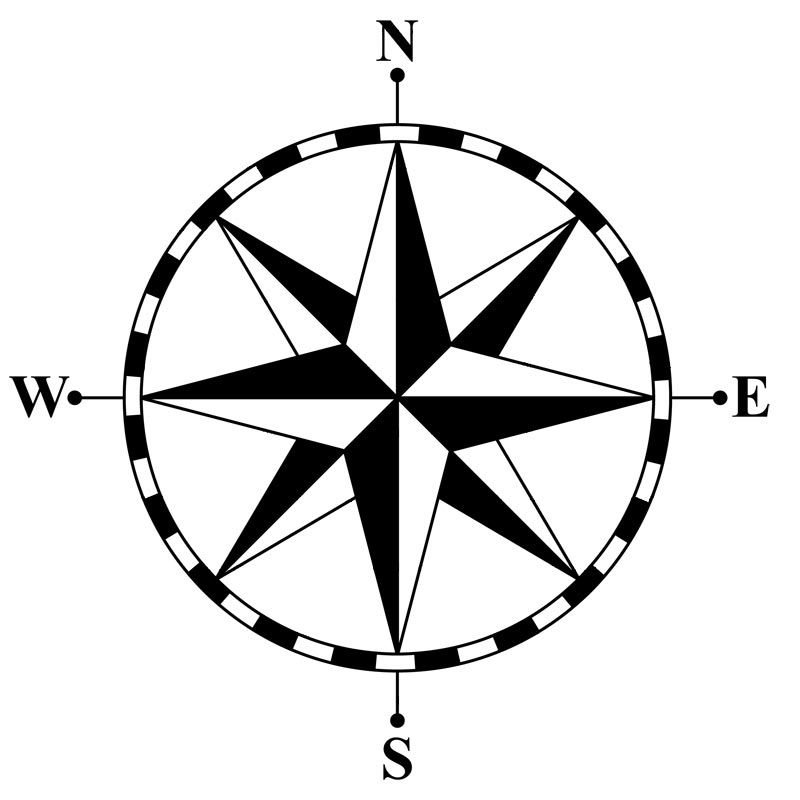}}  
  \end{overpic}
  \caption{Layout of HAWC WCDs and positions of the PMTs (PMTs not to scale).
  The conspicuous gap indicates the location of the counting house, which is centrally located to 
  minimize the cable length.}
  \label{fig:layout}
\end{figure}

  The HAWC detector is located in central Mexico at 18$^\circ$59'41"N 
97$^\circ$18'30.6"W and an elevation of 4100\,m a.s.l.
The instrument comprises 300 identical water Cherenkov detectors (WCDs) made from 5\,m high,
7.32\,m diameter commercial water storage tanks. Each tank contains a custom-made light-tight
bladder
to accommodate 190,000 liters of purified water. Four upward facing photomultiplier tubes (PMTs) are mounted at 
the bottom of each tank: a 10'' Hamamatsu R7081-HQE PMT positioned at the center and
three 8'' Hamamatsu R5912 PMTs which are positioned 
halfway between the tank center and rim. The central PMT has
roughly twice the sensitivity of the outer PMTs due to its superior quantum efficiency and its larger size.
The WCDs are filled to a depth of 4.5\,m, with 4.0\,m (more than 10 radiation lengths) of water
above the PMTs. This large depth guarantees that the electrons, positrons, and gammas in the air shower
are fully absorbed by the HAWC detector well above the PMT level, so that the detector itself acts 
as an electromagnetic (EM) calorimeter providing an accurate measurement of EM energy deposition.
High-energy electrons are detected via the Cherenkov light they produce in the water and
gamma rays are converted to electrons through pair production and Compton scattering. 
Muons are also detected.
They are more likely to be produced in air showers originating from hadronic
cosmic-ray interactions with the atmosphere and tend to have higher transverse
momentum producing large signals in the PMTs far from the air shower axis and thus
serve as useful tags for rejecting hadronic backgrounds.
The WCDs are arranged in a compact layout to maximize the density of the sensitive area, with 
about 60\% of the 22,000\,m$^2$ detector area instrumented. See Figure~\ref{fig:layout} for a diagram of the HAWC 
detector. 

Analog signals from the PMTs are transmitted by RG-59 coaxial cable to a central counting house.
The signals are shaped and discriminated at two voltage thresholds roughly corresponding to 1/4\,PE and
4\,PEs and the threshold crossing times (both rising and falling) are recorded using CAEN V1190A time-to-digital converters.
Individual signals that pass at least the low threshold are called hits.
The time-over-threshold is used to estimate the charge. The response of this system is roughly logarithmic, so that the 
readout has reasonable charge resolution over a very wide dynamic range, from a fraction of 1\,PE to  10,000\,PEs.
The timing resolution for large pulses is better than 1\,ns.  All channels are read out in 
real time with zero dead time and blocks of data are aggregated in a real-time computing farm. A trigger is 
generated when a sufficient number of PMTs record a hit within a 150\,ns window
(28 hits were required for most of the data used in this analysis, though other values were occasionally used earlier).
This results in a $\sim$20\,kHz trigger rate. Small events, with a number of hits close to the threshold value and which dominate the triggers, require a specific treatment
and are removed from the analysis presented here.
In the future their inclusion will significantly lower 
the energy threshold of HAWC. For sources with spectra that extend beyond 1\,TeV,
like the Crab Nebula, the sensitivity usually peaks above 5\,TeV (depending on
the source spectrum and declination)
and excluding the near-threshold events does not significantly
reduce the sensitivity.
Details of the event selection for the present analysis are presented in the next section.

For each triggered event, the parameters of the air shower, like the direction,
the size, and some gamma/hadron separation variables,
are extracted from the
recorded hit times and amplitudes, using a shower model developed through
the study of Monte Carlo simulations and optimized using observations
of the Crab Nebula \citep{crabpaper}.
The angular resolution of the HAWC instrument varies with the event size (number of hit
PMTs) and
ranges from $\sim$0.2\degree (68\% containment) for large events events hitting
almost all the PMTs to $\sim$1.0\degree for events near the 
analysis threshold.

Gamma-ray induced showers are generally compact and have a smooth 
lateral distribution around the shower core (the position where the shower axis
intersects the detector plane).
In contrast, hadronic background events tend to be broader, contain 
multiple or poorly defined cores, and include highly localized large signals from muons and 
hadrons at significant distance from the shower axis.
Selection cuts on shower morphology eliminate ${>}99\%$ of the hadronic
background in the large event size samples and at least $85\%$ of the background near the analysis 
threshold, while usually retaining more than 50\% of the gamma-ray induced signal events. 
Details of the data reconstruction,  and analysis, and the verification of the
sensitivity of the measurement 
will be presented in a future publication on the observation of the Crab Nebula
with the HAWC Observatory \citep{crabpaper}.

\section{Methodology}
  \label{sec:method}
  In this section we review the details of the dataset used in the analysis and
describe the event selection and the construction of unbiased maps of the viewable sky, which include estimates of
the cosmic-ray background rates. From the maps we compute a test statistic (TS) from the ratio
of the likelihood that a source is present and the null hypotheses that the observed
event population is due to background alone. We identify and localize sources from a 
list of local maxima in the TS maps with values greater than 25.
The procedure is applied to the map to identify pointlike sources as well as
sources with characteristic sizes 0.25\degree, 0.5\degree, 1\degree, and 2\degree.
Many sources, particularly the bright ones, will likely be detected in both the 
point-source and extended-source maps. We find that there are some extended regions of gamma-ray
emission that could either be interpreted as a single extended source or an ensemble of point sources.
Below we describe the method employed to detect point and extended sources, to estimate their positions, extents, and spectra
and finally discuss the principal sources of systematic uncertainty.

\subsection{Dataset}

  The results presented here are obtained using data taken between 2014-11-26 and
  2016-06-02.  During this period, $8.8 \times 10^{11}$ triggered events were
  recorded to disk.
  The full HAWC Observatory was inaugurated in 2015 March.
  During the construction phase prior to the inauguration, data were collected
  with a variable number of WCDs ranging from 250 to 300.
  Overall there was downtime of 40 days (7.2\%) during this 553 day period,
  for the most part related to power issues or scheduled shutdowns for construction or maintenance.
  In addition, 7 days of data (1.3\%) were removed based on requirements
  regarding the stability of the detector performance.
  The final livetime used for the analysis is 506.6\,days, corresponding to 92\% duty cycle.
  
  The data were reconstructed 
  and analyzed with \emph{Pass~4}, which includes improved calibrations, improved event
  reconstruction, and improvements in the  likelihood framework used for the map
  analysis.  The new event reconstruction benefits from a directional fit
  using an improved shower model, a new algorithm to separate gamma-ray and hadronic
  events, and a better electronics model.
  For comparison, our previous search for sources in the inner Galactic Plane
  which defined the 1HWC source list
  \citep{hawc111paper} was performed using 275~days of data taken with a detector
  consisting of about one third of the full HAWC array and using the \emph{Pass~1} analysis.
  This new pass, combined with the larger detector and longer exposure time,
  improves the sensitivity of the survey by about a factor of 5 with respect to
  the \emph{Pass~1} inner Galactic Plane search.

  \subsection{Event Selection\label{sec:reco}}
  
  Events are classified by size in nine analysis bins $\mathcal{B}$, presented in
  Table~\ref{tab:bins}, depending on the fraction $f_\mathrm{hit}$ of active PMTs in the
  detector that participate in the reconstruction of the air shower.
  We chose to define bins based on the 
  fraction of the detector hit, rather than the absolute number of PMTs, in order
  to obtain more stable results for the various detector configurations of
  active WCDs over time. 
  
  The selection cuts on the gamma/hadron separation variables are optimized for
  each bin using observations of the Crab Nebula \citep{crabpaper}.
  The point spread function (PSF) of the reconstructed events depends on the event size.
  In Table~\ref{tab:bins}, the $\psi_{68}$ column represents the 68\% containment angle of the PSF, for a
  source similar to the Crab Nebula.
  Large events have a better PSF, a better hadronic background
  rejection, and correspond to higher energy primary particles.
  The efficiency of the gamma/hadron separation cuts is indicated in the 
  $\epsilon_\gamma^\mathrm{MC}$ and $\epsilon_\mathrm{CR}^\mathrm{data}$
  columns, where the gamma efficiency has been estimated using Monte Carlo
  simulation of the detector and the hadron efficiency has been measured
  directly using cosmic ray data.
  The $\tilde{E}_\gamma^\mathrm{MC}$ column represents the median energy of the 
  simulated gamma-ray photons in this analysis bin for a source
  at a declination of 20\degree and for an energy spectrum $E^{-2.63}$ (Crab-Nebula-like source).
  Events in the same bin for a source with a harder spectrum or at larger
  declination will tend to have a larger energy on average.
    
  \begin{deluxetable}{ c c c c c c}
    \tablecaption{Properties of the nine analysis bins: bin number $\mathcal{B}$,
    event size $f_\mathrm{hit}$, 68\% PSF containment $\psi_{68}$, cut selection
    efficiency for gammas $\epsilon_\gamma^\mathrm{MC}$ and cosmic rays
    $\epsilon_\mathrm{CR}^\mathrm{data}$, and median energy for a reference
    source of spectral index $-2.63$ at a declination of 20\degree
    $\tilde{E}_\gamma^\mathrm{MC}$.
    \label{tab:bins}}
    \tablehead{
      \colhead{$\mathcal{B}$} &
      \colhead{$f_\mathrm{hit}$} &
      \colhead{$\psi_{68}$} &
      \colhead{$\epsilon_\gamma^\mathrm{MC}$} &
      \colhead{$\epsilon_\mathrm{CR}^\mathrm{data}$} &
      \colhead{$\tilde{E}_\gamma^\mathrm{MC}$} \\
      \colhead{} &
      \colhead{(\%)} &
      \colhead{(\degree)} &
      \colhead{(\%)} &
      \colhead{(\%)} &
      \colhead{(TeV)}
    }
    \startdata
      1 &  6.7 --  10.5 & 1.03 & 70 &   15 & 0.7 \\
      2 & 10.5 --  16.2 & 0.69 & 75 &   10 & 1.1 \\
      3 & 16.2 --  24.7 & 0.50 & 74 &  5.3 & 1.8 \\
      4 & 24.7 --  35.6 & 0.39 & 51 &  1.3 & 3.5 \\
      5 & 35.6 --  48.5 & 0.30 & 50 & 0.55 & 5.6 \\
      6 & 48.5 --  61.8 & 0.28 & 35 & 0.21 &  12 \\
      7 & 61.8 --  74.0 & 0.22 & 63 & 0.24 &  15 \\
      8 & 74.0 --  84.0 & 0.20 & 63 & 0.13 &  21 \\
      9 & 84.0 -- 100.0 & 0.17 & 70 & 0.20 &  51 \\
    \enddata
  \end{deluxetable}
  
  \subsection{Event and Background Maps}
  
    After reconstruction, event and background maps are generated.
    The event maps are simply histograms of the arrival direction of the
    reconstructed events, in the equatorial coordinate system.
    The background maps are computed using a method developed for the Milagro experiment
    known as direct integration \citep{directintegration}.
    It is used to fit the 
    isotropic distribution of events that pass the gamma-ray event
    selection, while accounting for the asymmetric detector angular response and
    varying all-sky rate.
    As strong gamma-ray sources would bias the background estimate, some regions
    are excluded from the computation.
    These regions cover the Crab, the two Markarians, the Geminga region and, a
    region $\pm3$\degree around the inner Galactic Plane.
    Nine event maps and nine background maps are generated, for the nine analysis bins.

    The maps are produced using a HEALPix pixelization scheme \citep{healpix},
    where the sphere is divided in 12 equal area base pixels, each of which is
    subdivided into a grid of $N_\mathrm{side} \times N_\mathrm{side}$.
    For the present analysis, maps were initially done using
    $N_\mathrm{side}=1024$ for a mean spacing between pixel centers of less than 0.06\degree, which is small
    compared to the typical PSF of the reconstructed events as shown on Table~\ref{tab:bins}.

  \subsection{Source Hypothesis Testing}
  
    The maximum likelihood analysis framework presented in \citet{lifficrc} is
    used to analyze the maps.  The test statistic is defined using the likelihood ratio,
    \begin{equation} \label{eq:likelihood}
      TS=2\ \textrm{ln}\frac{\mathcal{L}^\mathrm{max}(\textrm{Source Model})}{\mathcal{L}(\textrm{Null Model})}\,,
    \end{equation}
    to compare a source model hypothesis with a null hypothesis.
    The likelihood of a model $\mathcal{L}(\textrm{Model})$ is obtained by
    comparing the observed event counts with the expected counts, for all the
    pixels in a region of interest, and for all nine analysis bins.
    
    For the null model, the expected counts are simply given by the background
    maps derived from data.
    For the source model, the expected counts correspond to the same background
    plus a signal contribution from the source derived from simulation.
    We assume a source model characterized by:
    \begin{itemize}
      \item a point source or a uniform disk of fixed radius and
      \item a power law energy spectrum.
    \end{itemize}
    The signal contribution is derived from the source characteristics and the
    detector response from simulation (expected counts for the spectrum
    and PSF, both functions of the analysis bin and the declination).

    The TS is maximized with respect to the free parameters of the source model.
    This approach is used both to search for sources (with a TS threshold)
    and to measure the characteristics of said sources as a result of
    the maximization.
    
    \begin{figure}
      \centering
      \includegraphics[trim={0cm 0cm 0cm 0cm},clip,width=\columnwidth]{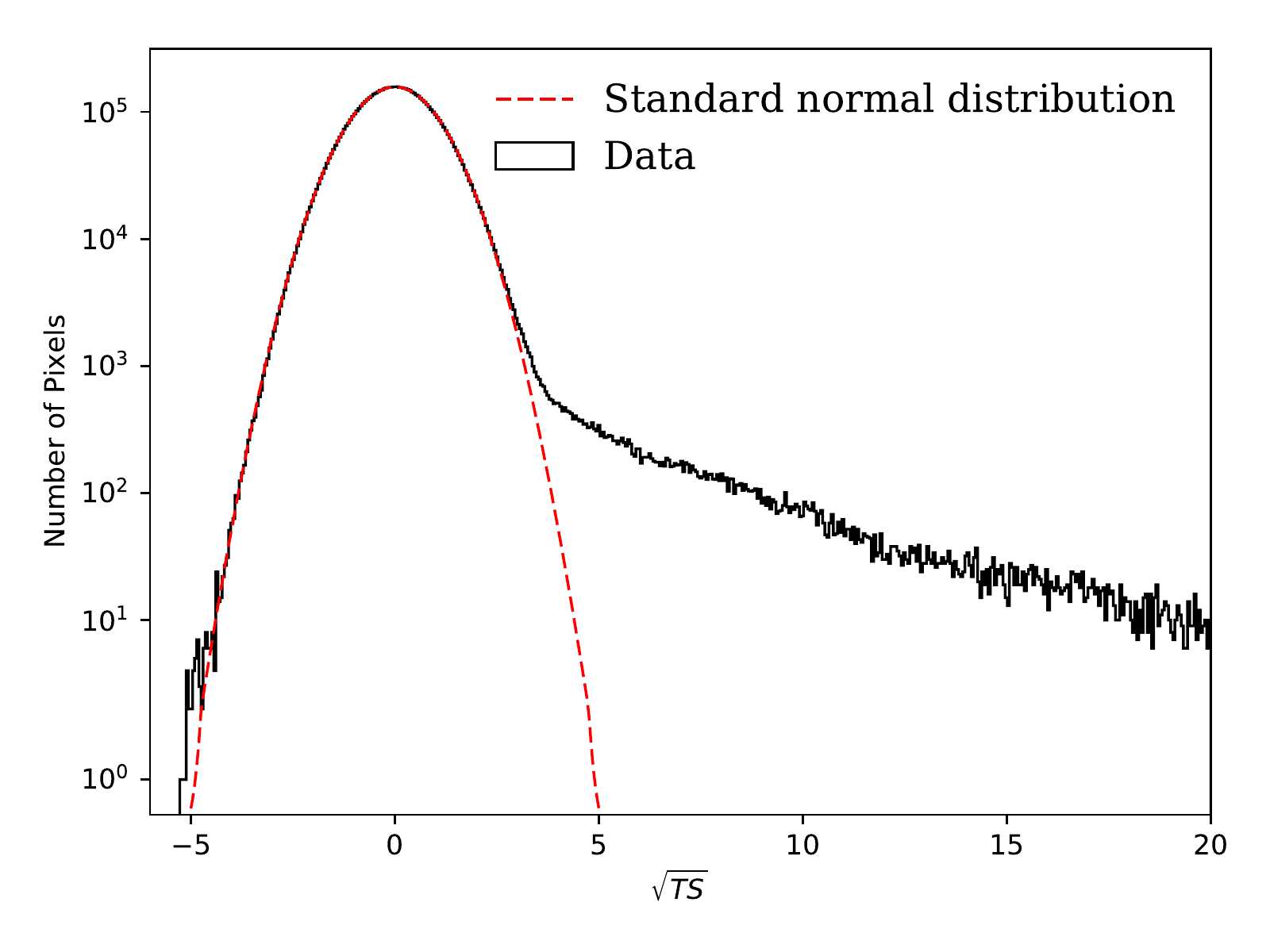}
      \caption{Test statistic distribution of the point source search (black)
      and standard normal distribution (red).}
      \label{fig:histsigma}
    \end{figure}
    
    We make a TS map by moving the location of the hypothetical source across the
    possible locations in the sky.
    In the following searches the source flux is the only free
    parameter of the model while the extent and spectral index are fixed.
    The source and null model are nested; hence by Wilks' Theorem the TS is
    distributed as $\chi^2$ with one degree of freedom if the statistics are
    sufficiently large.
    Consequently, the pre-trial significance, conventionally reported as
    standard deviations (sigmas), is obtained by taking the square root of the
    test statistic, \sts (here and after, what we denote \sts actually
    corresponds to $\textrm{sign}(TS)\,\sqrt{|TS|}$).
    Figure~\ref{fig:histsigma} shows the distribution of \sts across
    the sky for the point source search, as well as a standard normal distribution
    scaled by the number of pixels.
    For values lower than ${\sim}3$, the \sts is well reproduced by the normal
    distribution, whereas at greater values a large excess can be seen due to
    the presence of sources in the sky.
    
\subsection{Catalog Construction}\label{sec:method-catalog}

    In order to take advantage of HAWC's sensitivity to both pointlike and
    extended sources, multiple searches are conducted assuming either point or
    extended sources.
    The TS maps used for the search are computed using a source model
    consisting of a single test source with a fixed
    geometry (point source or uniform disk of fixed radius) and an energy
    spectrum consisting of a power law of fixed index,
    \begin{equation} \label{eq:spectrum}
      dN/dE = F_0 (E/E_0)^\alpha\,,
    \end{equation}
    where $E_0$ is a reference energy, $F_0$ is
    the differential flux at $E_0$ and $\alpha$ is the spectral index.

    For the known TeVCat sources that can be considered pointlike given the
    angular resolution of the HAWC instrument (i.e. the TeVCat extent is of the
    order of the PSF size or smaller), the spectral indices measured by HAWC vary around
    $-2.7$, from approximately $-3.1$ to $-2.5$, and are typically softer than
    the indices listed in TeVCat.
    This can be explained if the sources soften
    or cut off at the energies observed by HAWC.
    On the other hand, the Geminga PWN, which was first observed at TeV energies by the
    Milagro collaboration \citep{Abdo:2009ku}, is detected by HAWC with an
    extent of about 2\degree and a hard spectral index around $-2$.
    To account for the range of source extents and spectra observed with
    HAWC, four different maps were used to build the catalog, testing
    various source hypotheses. In order to limit computing time, the resolutions
    of the maps are adapted to the characteristic dimension of the hypothetical
    source, without significantly affecting the results:
    \begin{enumerate}
      \item A point source map of index $-2.7$ (HEALPix map resolution $N_\mathrm{side}=1024$ or 0.06\degree per pixel).
      \item An extended source map of radius 0.5\degree and index $-2.0$ ($N_\mathrm{side}=512$ or 0.1\degree per pixel).
      \item An extended source map of radius 1.0\degree and index $-2.0$ ($N_\mathrm{side}=256$ or 0.2\degree per pixel).
      \item An extended source map of radius 2.0\degree and index $-2.0$ ($N_\mathrm{side}=256$ or 0.2\degree per pixel).
    \end{enumerate}
    
    When building the catalog, the priority is given to the point source search,
    then the extended searches ordered by increasing radius.
    This limits possible source contamination when multiple nearby sources
    are added together.
    However, a strong extended source may be found in the point source
    search, possibly multiple times (see e.g. Geminga below), as well as in the extended search.
    Hence, the exact search in which a source is first tagged is not a perfect
    indication of the source extent.
    More robust morphology studies will be performed in a future analysis and are
    beyond the scope of this catalog paper.
    
    \label{sec:search}
    To select the sources in the maps, all local maxima with $\mathrm{TS}>25$ are flagged.
    In some regions, multiple local maxima are found very near each other.
    We define as primary sources all local maxima that are separated
    from neighboring local maxima of higher significance by a valley of $\Delta(\sqrt{\mathrm{TS}})>2$.
    We also define and include secondary sources when
    $1<\Delta(\sqrt{\mathrm{TS}})<2$.
    These sources are marked with an asterisk (*).
  
    The final catalog comprises the sources of the point source search plus the
    sources of the extended searches, ordered by increasing radius, if their locations are
    more than 2\degree away from any hotspot with TS greater than 25 in the
    previous searches.

  \subsection{False Positive Expectation\label{sec:trials}} 

    When selecting the sources in the map, a background fluctuation can
    sometimes mimic a source and fulfill the selection criteria.
    To estimate this possible contamination, the search was run on randomized
    background maps.
    Events maps are generated for each of the nine analysis bins, and then the
    full search strategy as for the data map is employed, including point and
    extended source searches, as detailed on
    Section~\ref{sec:method-catalog}.
    This complete procedure was run with 20 sets of simulated maps.
    In 11 cases, no sources were flagged. In 9 cases, one source was flagged.
    In total, out of the 20 full searches performed over the entire sky, 9
    sources were flagged, so the predicted number of
    background fluctuations passing the $\mathrm{TS}>25$ criterion is about $9/20=0.45$.
    Therefore, the predicted number of false positive in the catalog is about 0.5.
    These possible fluctuations are typically close to the threshold value
    $\mathrm{TS}=25$ and are usually out of the Galactic Plane, as it only
    represents a small fraction of the visible sky.
  
  \subsection{Source Position, Extent, and Energy Spectrum\label{sec:parameter-estimate}}
  
    The source positions reported in this catalog correspond to
    the first search in which they appear, as presented in Section~\ref{sec:search}.
    The statistical uncertainty of the position is defined as the maximum
    distance between the center and the 1-sigma contour
    obtained from the TS map.
    
    After the search, a residual map is generated and
    halo-like structures are visible around several sources modeled as point
    sources.
    This halo is used to define a tentative source radius for the
    secondary source model when fitting the energy spectrum (results presented
    in Table~\ref{tab:fluxes} of the next section).
    This radius should not be regarded as a definite measurement of the source
    extent but can nonetheless
    provide useful information on how much the spectrum measurement depends on
    the source region definition.
    When this new source region definition is a good representation of the
    actual source, the newly fitted spectrum should better correspond to the
    source spectrum, however as it corresponds to a larger region it is more
    subject to contamination from other sources or possibly diffuse emission.
    Additionally, for some complex regions, or regions for
    which independent analyses are performed, the whole region is
    fit, explicitly including multiple sources, as an estimate of the
    total flux of the region. Such regions are discussed in
    Section~\ref{sec:discussion}.
  
    Once the source location and size are defined, the source spectrum is fit
    using a power law (Equation~\ref{eq:spectrum}).
    For the range of declinations considered, the reference energy of 7\,TeV
    minimizes the correlation between the index and normalization, 
    energy which corresponds to the region of maximum sensitivity
    (cf. Figure~\ref{fig:sensitivity}, right).
    We report the differential flux at 7\,TeV ($F_7$), the index $\alpha$, and
    the  statistical uncertainties on both parameters in Table~\ref{tab:fluxes}.

\subsection{Diffuse Galactic emission\label{sec:diffuse}}

  At GeV energies, diffuse emission resulting from the interaction of
  cosmic rays with matter and photons is the dominant component of the gamma-ray sky.
  This diffuse emission has a steeper spectrum than galactic gamma-ray sources and
  as a result the TeV sky is source dominated.
  The Milagro and \HESS experiments measured the TeV diffuse emission in
  \citet{0004-637X-688-2-1078} and \citet{2014PhRvD..90l2007A}.
  Both measured a higher flux than predicted -- by the numerical cosmic-ray
  propagation code GALPROP \citep{strong2007cosmic} for Milagro\footnote{The conventional
  GALPROP version here, since the optimized version was derived to
  fit the EGRET excess which was latter refuted by \LAT.}, and a hadronic model for \HESS --, likely due to
  unresolved sources.
  A diffuse emission is not included in the likelihood model used in the present analysis.
  We are concerned that sources identified by this analysis may have a significant
  underlying diffuse component, or in extreme cases arise from background
  fluctuations in a continuous region of diffuse emission.
  To estimate the maximum possible contribution of
  the diffuse emission to the spectrum
  measurement, we simulate a uniform flux with a normalization corresponding to
  the peak flux value of the hadronic model reported by \HESS 
  ($1 \times 10^{-9}$\,TeV$^{-1}$\,cm$^{-2}$\,s$^{-1}$\,sr$^{-1}$ at 1 TeV) and a
  spectral index of $-2.7$.
  We estimate that, for the low latitude sources near the detection threshold
  (where the diffuse contribution will be the largest),
  the diffuse emission can contribute to $<$30\% of the fluxes measured with the
  point source hypothesis.

  As an alternative method of estimating the contribution from Galactic Diffuse
  emission, we can use a region of the Galactic Plane with no detected
  sources to derive a conservative upper limit on this contribution.
  As with the analyses by HESS and Milagro mentioned above, this approach will
  naturally overestimate the diffuse component since it includes unresolved sources.
  We use the region with longitude $l$ between 56\degree and 64\degree and latitude
  $|b|<0.5$\degree, which does not contain detected sources.
  The median differential flux at 7\,TeV measured in this region with the point
  source model is $2.1 \times 10^{-15}$\dfu.
  This small excess over a large region indicates the presence of either the
  Galactic diffuse emission, some unresolved sources, or more likely a combination
  of both.
  We use it as an upper limit to estimate the impact of the diffuse on the flux of the
  sources measured in the plane near $l=60$\degree.
  We extrapolate to lower latitudes using the shape of the longitudinal
  profile of the diffuse emission from GALPROP in \citet{0004-637X-688-2-1078}.
  We find that in this approach the diffuse emission can contribute up to 60\% of
  the flux measurement of the weak, low-latitude sources (TS close to
  25), that have longitudes between 34\degree and 50\degree.
  For $l>50$\degree the modeled diffuse emission is lower, and for $l<34$\degree
  all the detected sources have higher fluxes and they are not impacted
  significantly by the diffuse emission.
  The sources for which this conservative estimate is above 30\% of the measured
  point source flux at 7\,TeV are 2HWC J1852+013*, 2HWC J1902+048*, 2HWC J1907+084*,
  2HWC J1914+117*, 2HWC J1921+131, and 2HWC J1922+140; as defined and discussed
  in Sections \ref{sec:results} and \ref{sec:discussion}.
  In the likely case in which part or most of the flux measured in the
  $l=[56^\circ,\,64^\circ]$ region indeed contains unresolved sources, the
  diffuse flux is lesser and so is its contribution of the flux reported on this
  catalog.

  Future dedicated analysis of the HAWC data will allow to better constrain the
  Galactic diffuse emission.

\subsection{Systematic Uncertainties\label{sec:systematics}}
    
    The absolute pointing of the HAWC Observatory is initially determined using a careful survey of the
    WCDs and PMTs and then refined using the observed position of the Crab Nebula.
    The positions of Markarian~421 and
    Markarian~501 are observed by HAWC within 0.05\degree of their known locations
    after the pointing calibration.
    Additional studies based on the observation of the Crab Nebula when it is
    farther from zenith showed that absolute pointing is still better than
    0.1\degree up to a zenith angle of 45\degree, which covers the full
    declination range considered in the present study.
    Therefore the systematic uncertainty on the absolute pointing of
    the catalog is quoted as 0.1\degree.
    
    For isolated point sources, the systematic uncertainties on the spectrum measurement are estimated to be
    $\pm 50 \%$ for the overall flux and $\pm 0.2$ for the spectral index \citep{crabpaper}.
    In the present analysis, no detailed morphology study is performed.
    However, there is a correlation between the assumed source size and
    the measured spectrum.
    Simulation studies show that for isolated sources the unknown extent can induce
    an additional systematic uncertainty on the spectral index measurement of up to 0.3.
    
    As we test the presence of a single source at a time without modeling
    the other sources, the likelihood computation may be impacted by events from a
    neighboring source.
    This is true in particular for the lower energy events where the PSF is
    wider.
    By adding events to the single hypothesized source, this contamination
    can increase the measured flux and make the spectral index softer.
    In the case of two identical point sources located 1\degree apart, the flux
    measurement, assuming a known spectral index, is increased by 20\% to 30\%,
    depending on the declination.
    When fitting the index as well, the index can change by up to 0.1 and the
    measured flux is changed by about 20\% to 40\%.
    This confusion is considered a systematic uncertainty of the present
    analysis and tends to be
    larger in the very populated regions of the sky with high source population.

\section{Results}
  \label{sec:results}
  We present the result of the search, the 2HWC catalog.
A total of 39 sources are found\footnote{Geminga is flagged twice but
only counted as one here.}, 4 of which are detected with
the extended search procedure only.
As discussed in Section~\ref{sec:trials}, the predicted number of background
fluctuations passing the selection criteria is about 0.5.
Out of these 39 sources, 16 are more than a degree away from known TeV
sources listed in TeVCat.

\subsection{HAWC Performance}\label{sec:performance}

  \begin{figure*}
    \centering
    \includegraphics[trim={0cm 0cm 1.3cm 0.8cm},clip,width=0.49\textwidth]{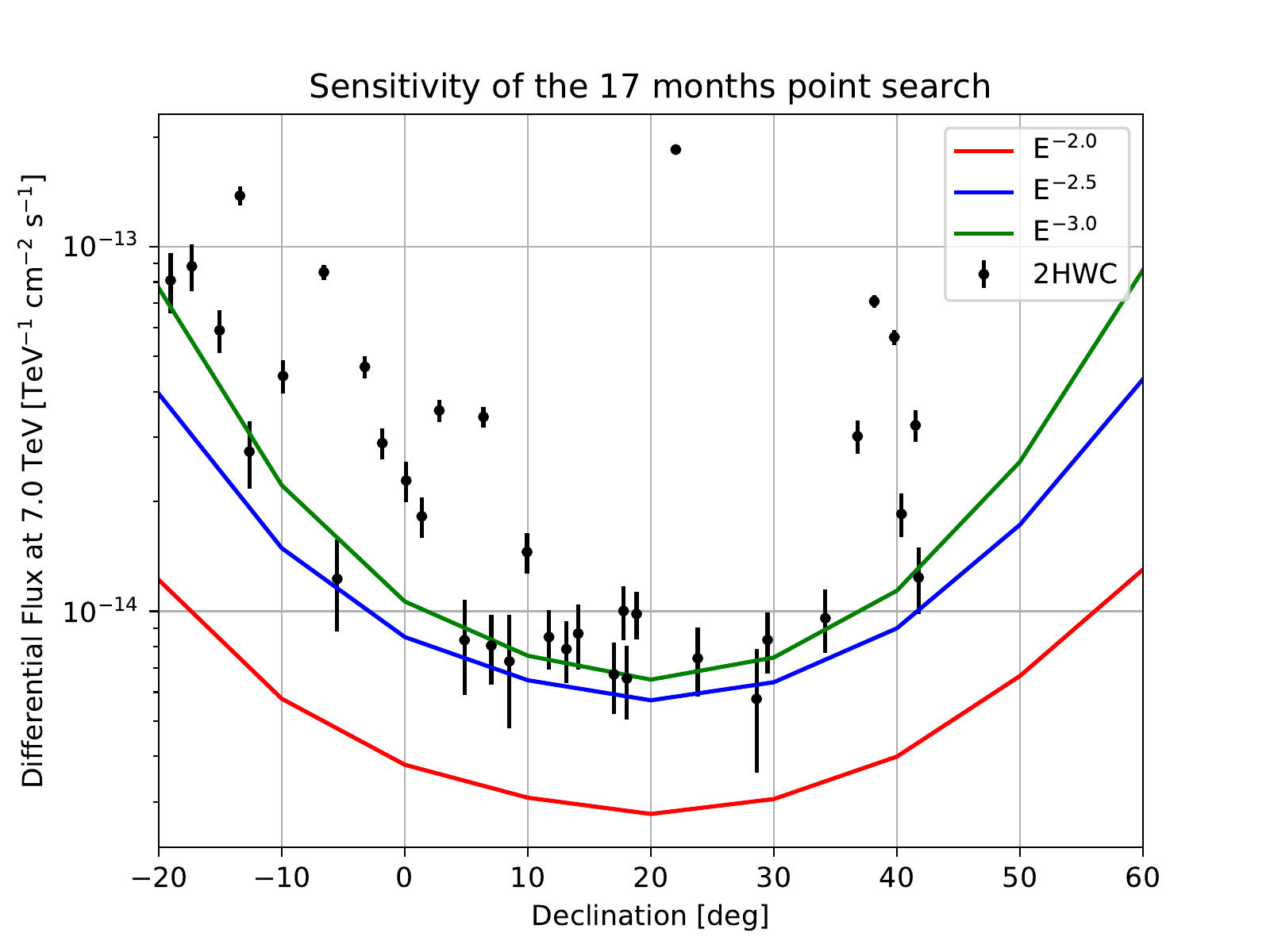}
    \includegraphics[trim={0cm 0cm 1.3cm 0.8cm},clip,width=0.49\textwidth]{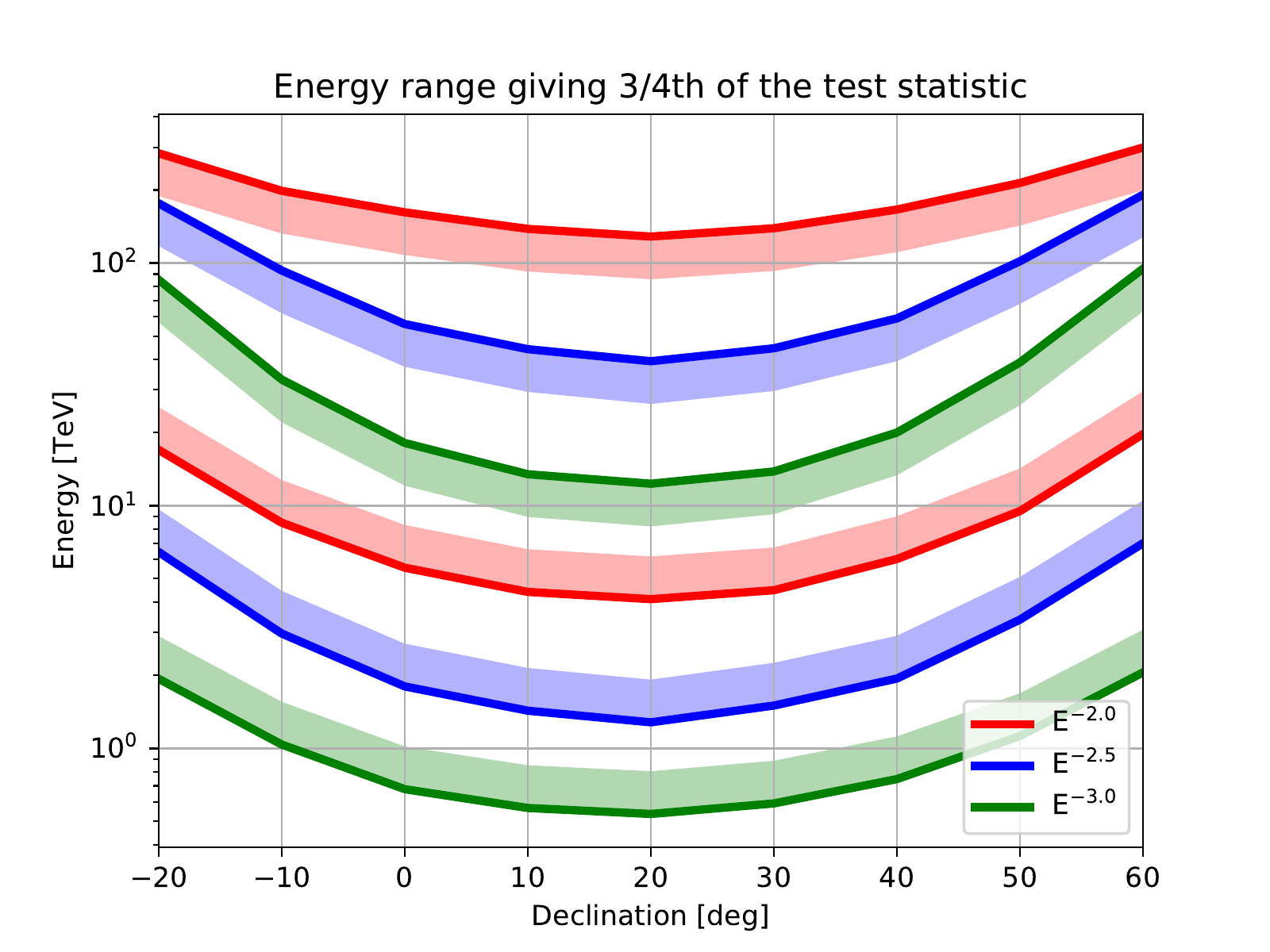}
    \caption{\emph{Left}: Sensitivity of the point source search for three
             spectral hypotheses, as a function of declination.
             We show the flux required to give a central expectation of
             5$\sigma$, for the present analysis.
             The differential fluxes of the sources detected in the point source search are also
             shown with their statistical uncertainties.
             \emph{Right}: Upper and lower ends of the energy range
             contributing to the central 3 quarters of the test
             statistic of the point source search, see text.}
    \label{fig:sensitivity}
  \end{figure*}
  
  Due to the development of air showers in the atmosphere, HAWC's sensitivity as
  well as energy response varies with the source declination.
  The sensitivity of the point source search is
  represented in Figure~\ref{fig:sensitivity}, left.
  The curves correspond to the flux that gives a central expectation of a
  5$\sigma$ signal for a point source with a power law flux of index $-2.0$,
  $-2.5$, and $-3.0$.
  The maximum sensitivity is obtained for sources transiting at the zenith of HAWC, i.e.
  whose declinations are close to 19\degree.
  The sources found in the point source search are also represented here: the
  measured flux and statistical uncertainty are shown at the corresponding
  declination.

  The energy range that contributes to
  most of the test statistic in the point source search, derived from
  simulation, is represented in Figure~\ref{fig:sensitivity}, right.
  More precisely, assuming a given spectral model, we show the energy range
  as the energy defining the central 75\% of the contribution to
  the test statistic.
  Three spectral models are represented: power laws of index $-2.0$,
  $-2.5$, and $-3.0$.
  For a given spectral model, the energy range that contributes most of the
  test statistic shifts to lower values for sources transiting overhead than for sources whose
  declinations are far from 19\degree.
  
\subsection{Maps}
\begin{figure*}
    \includegraphics[trim={0.5cm 0cm 0.5cm 2cm},clip,width=\textwidth]{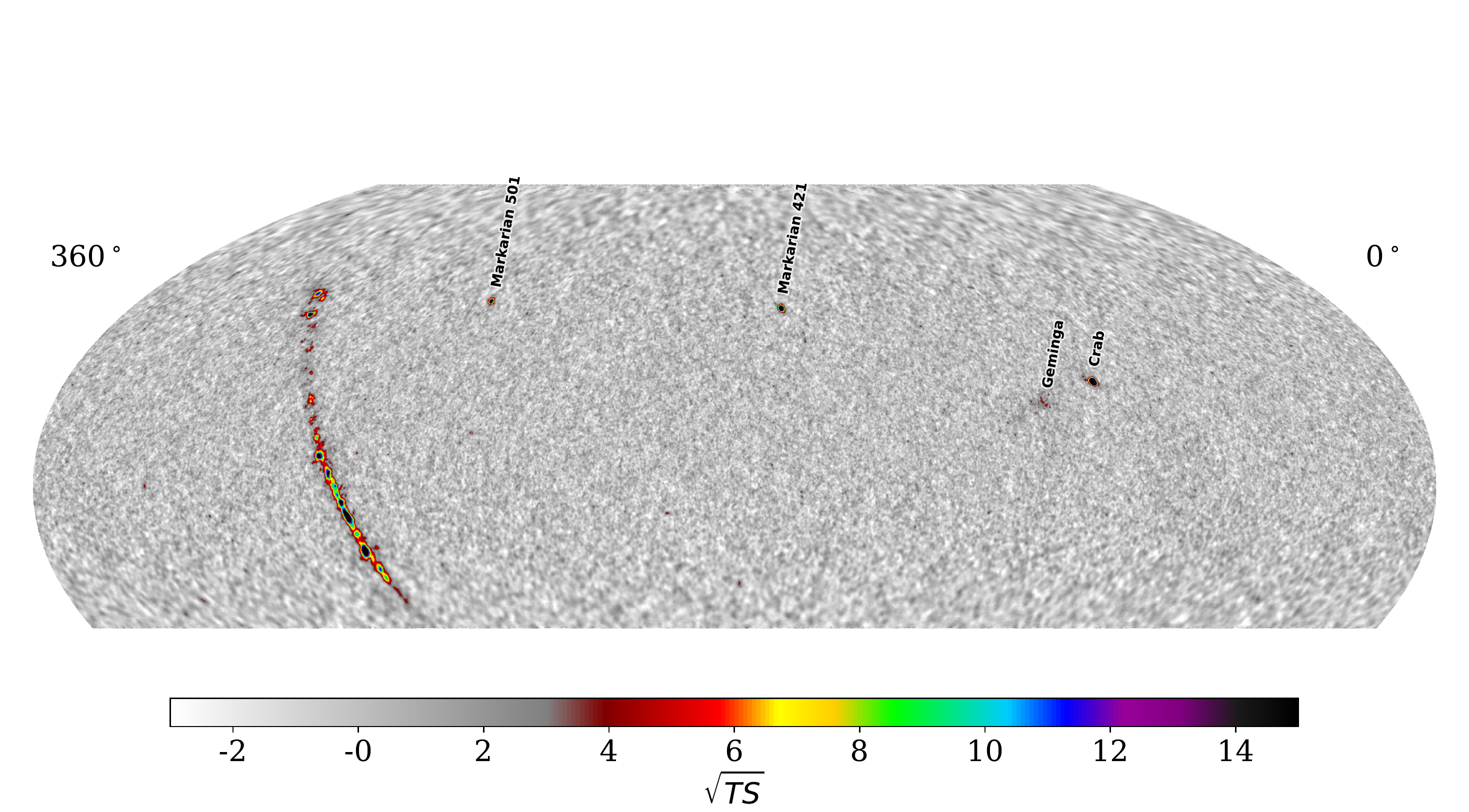}
    \caption{Equatorial full-sky TS map, for a point source hypothesis with a
    spectral index of $-2.7$.}
    \label{fig:equ-ps}
  \end{figure*}

  The test-statistic map derived from the all-sky search for point sources with index
  $-2.7$ is presented in equatorial coordinates in Figure~\ref{fig:equ-ps}.
  The inner Galactic Plane is clearly visible.
  In the outer Galactic Plane, the Crab and Geminga are visible.
  Outside of the Galactic Plane, Markarian 421 and Markarian 501 stand out.
  
\begin{figure*}
    \centering
    \includegraphics[trim={0cm 0cm 0cm 0cm},clip,height=0.32\textwidth]{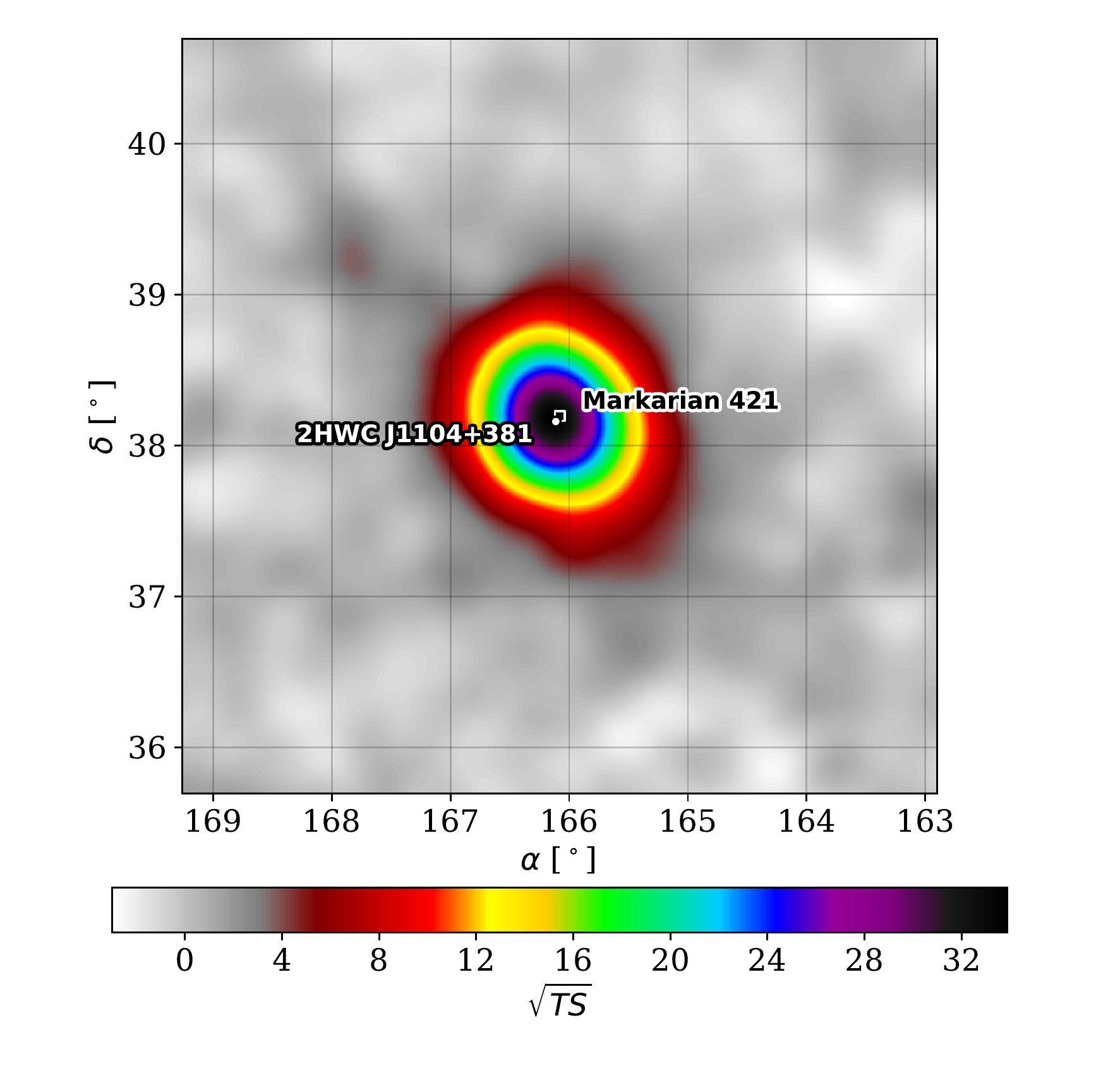}
    \includegraphics[trim={0cm 0cm 0cm 0cm},clip,height=0.32\textwidth]{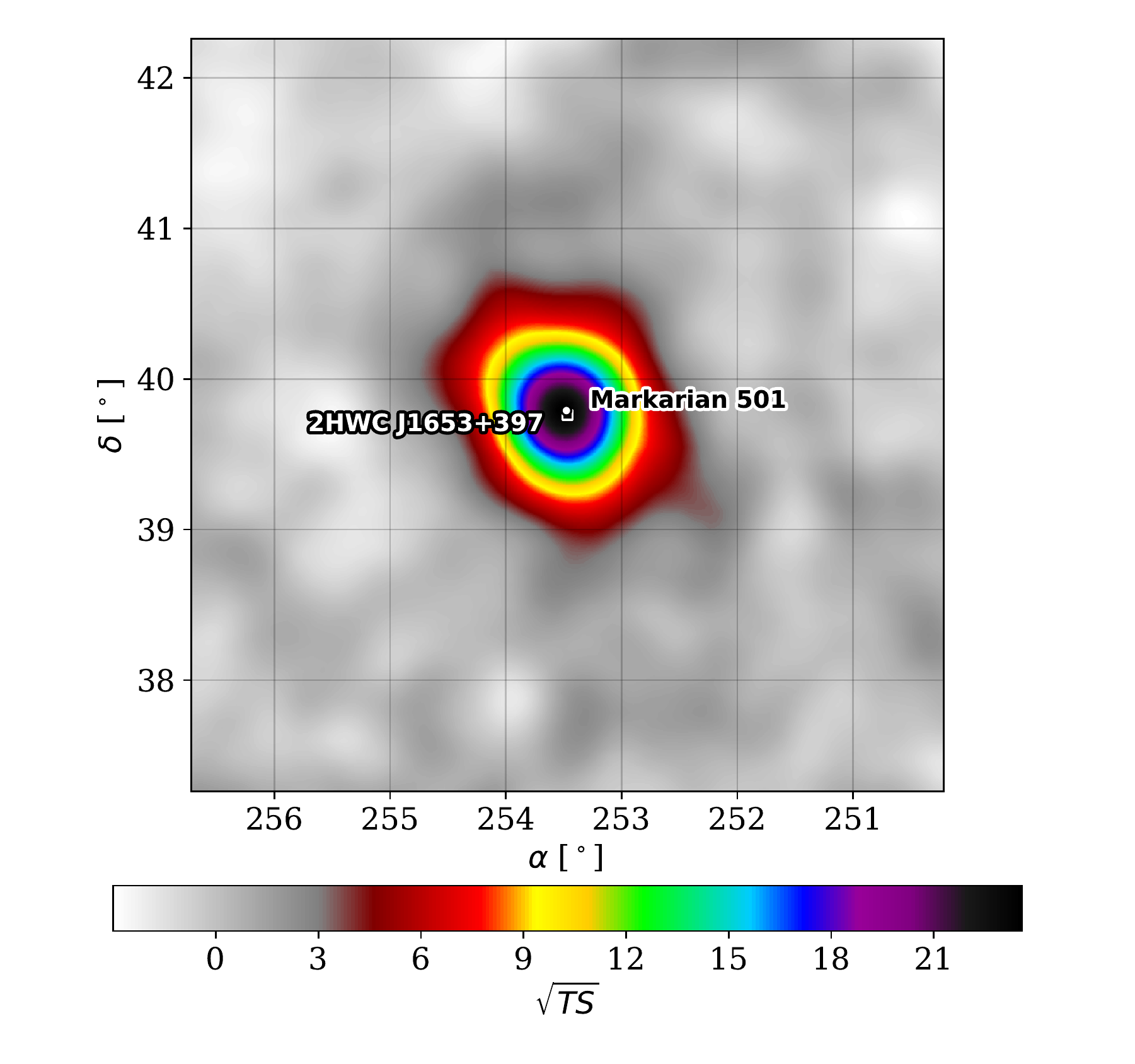}
    \includegraphics[trim={0cm 0cm 0cm 0cm},clip,height=0.32\textwidth]{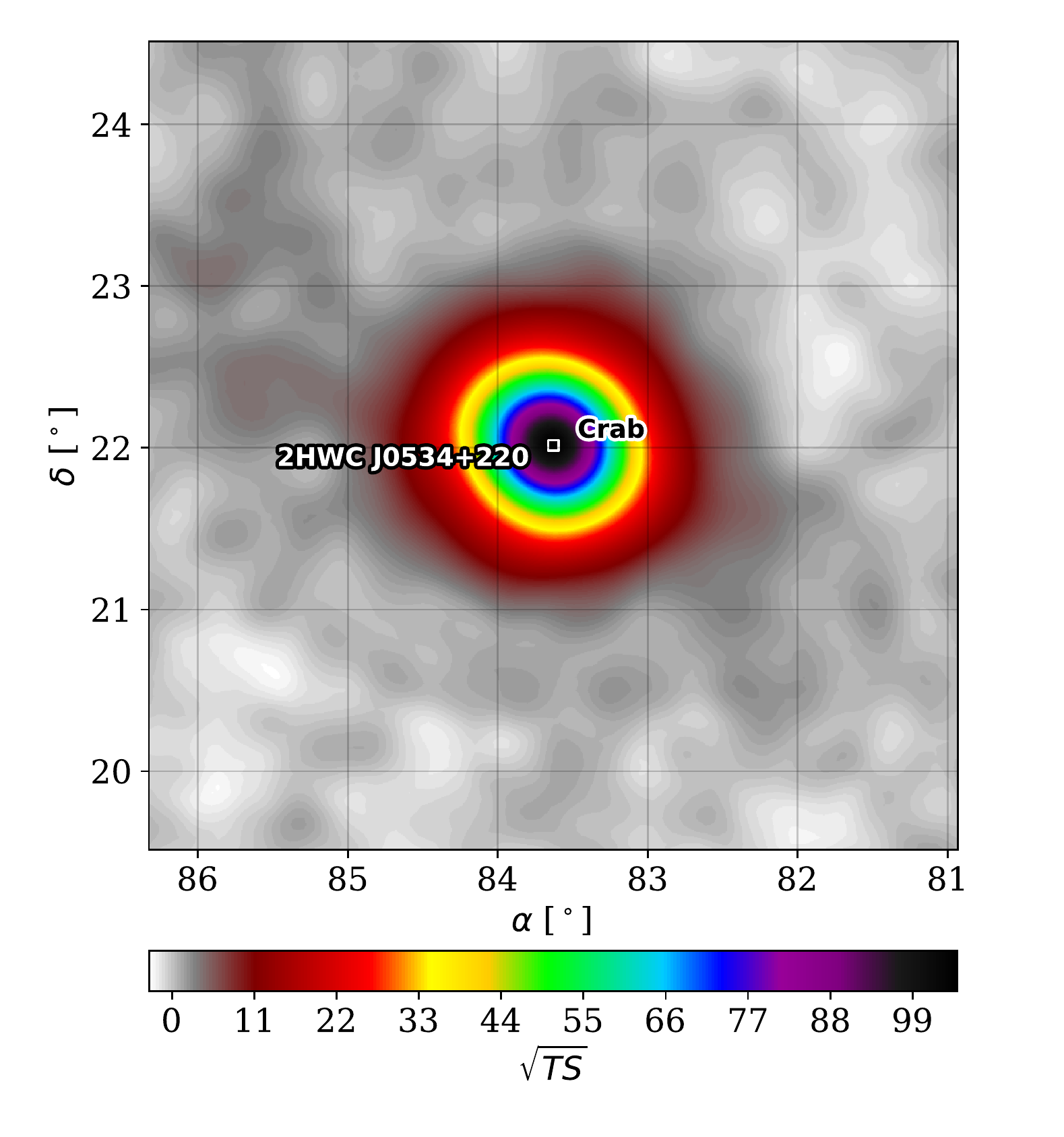}
    \caption{Regions around Markarian 421, Markarian 501, and the Crab Nebula: Equatorial TS
    maps, for a point source hypothesis with a spectral index of $-2.7$.
    In this figure and the followings, the 2HWC sources are represented by white
    circles and labels below the circle; whereas the source listed in TeVCat
    are represented with black squares and labels above the square symbol.}
    \label{fig:crab}
  \end{figure*}

\begin{figure*}
    \centering
    \includegraphics[trim={0cm 0cm 0cm 0cm},clip,width=0.4\textwidth]{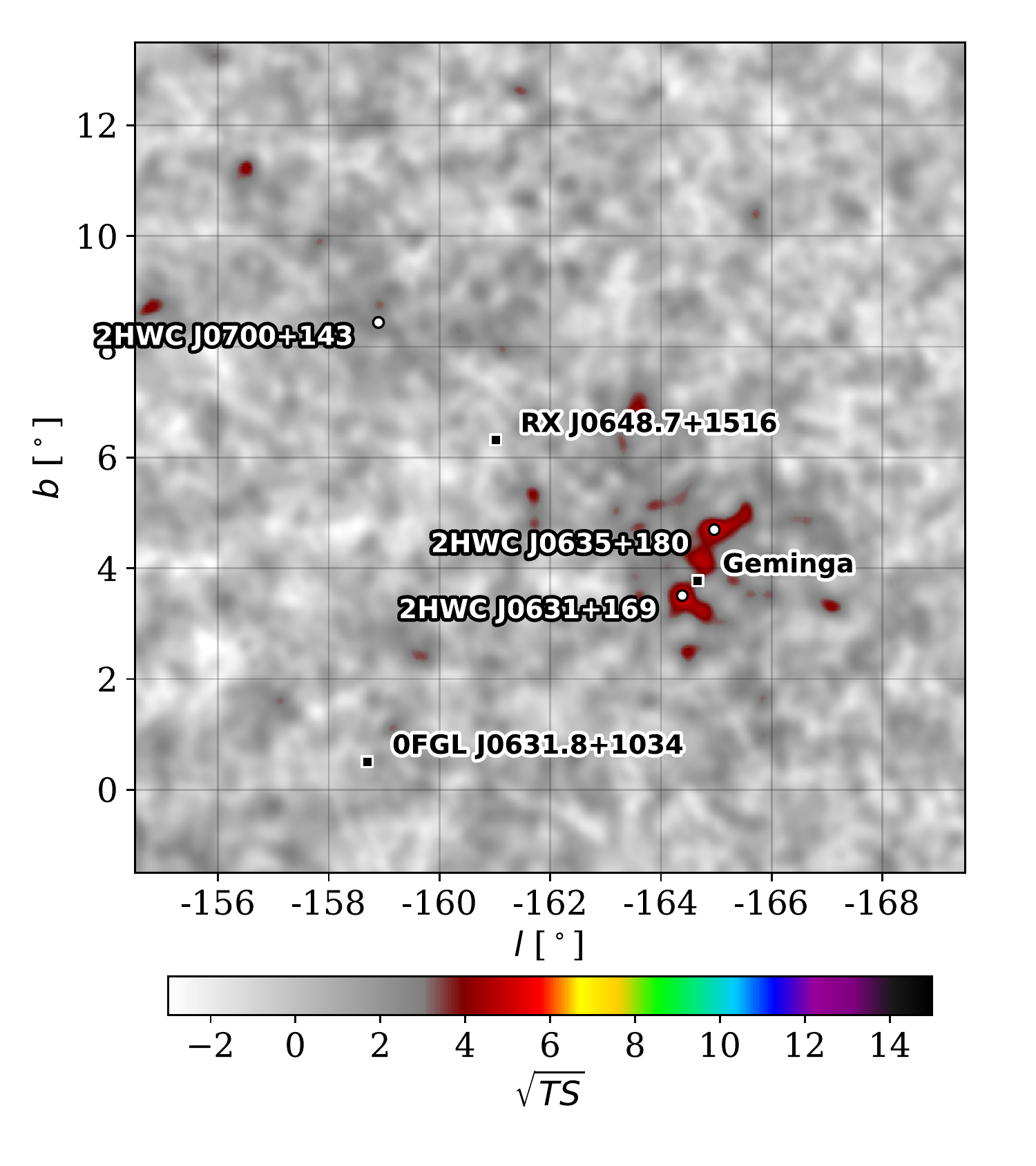}
    \includegraphics[trim={0cm 0cm 0cm 0cm},clip,width=0.4\textwidth]{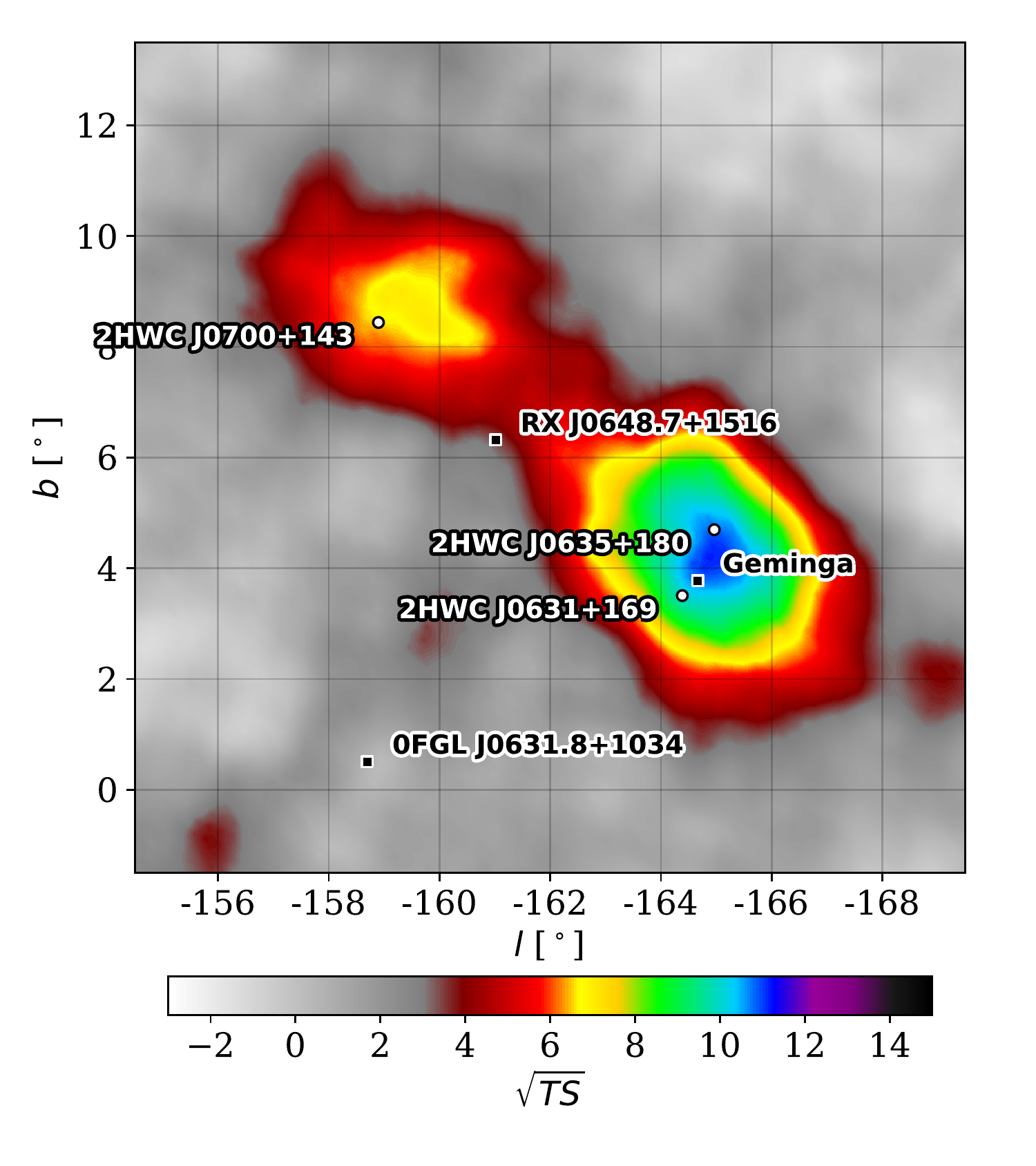}
    \caption{Region around Geminga, in Galactic coordinates. Left: TS map for
             a point source hypothesis with a spectral index of $-2.7$.
             Right: TS map for an extended source hypothesis represented by a
             disk of radius of 2.0 degrees with a spectral index of $-2.0$.
             }
    \label{fig:cgg}
  \end{figure*}

\begin{figure*}
    \centering
    \includegraphics[trim={0cm 0cm 0cm 0cm},clip,height=0.32\textwidth]{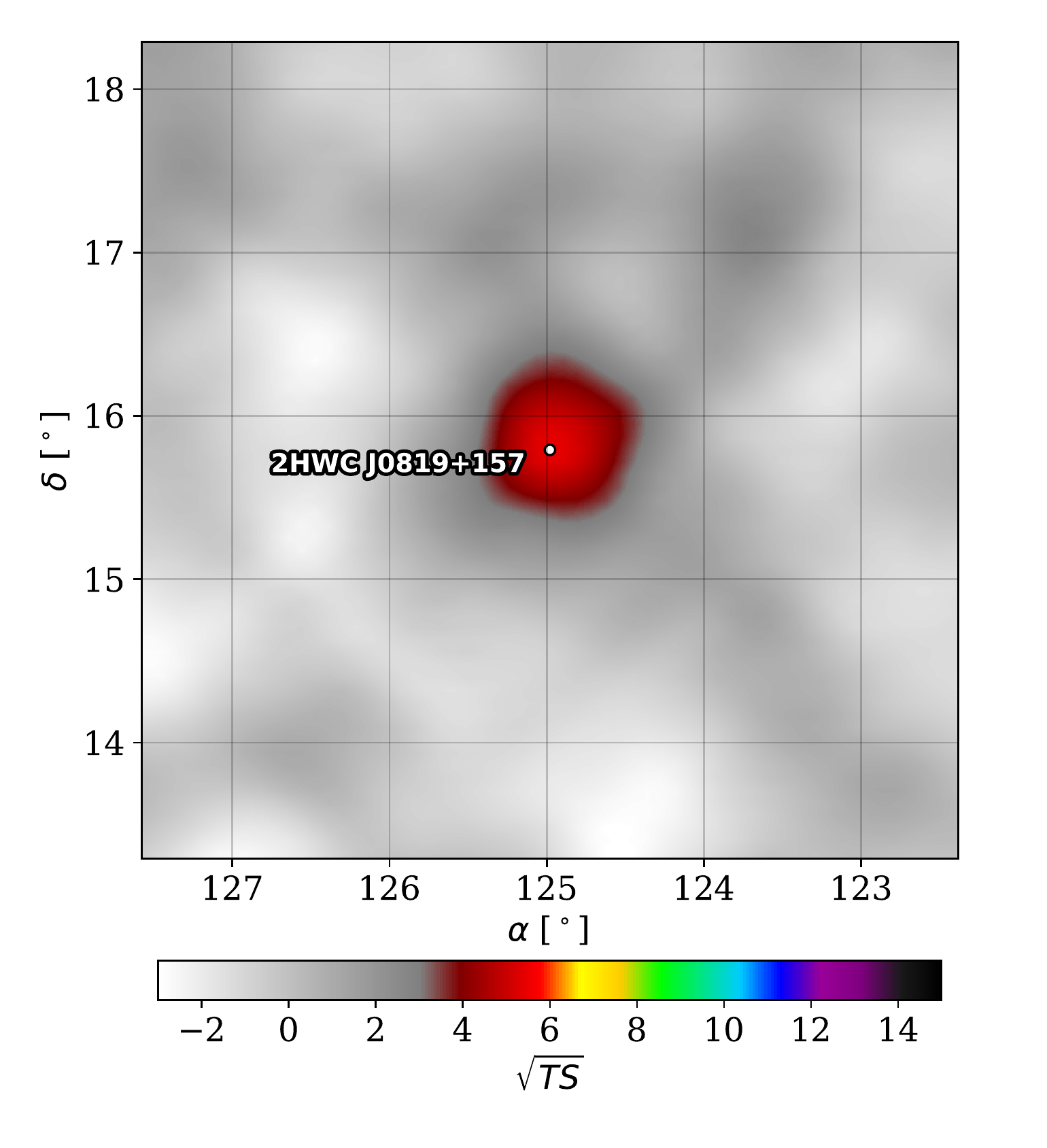}
    \includegraphics[trim={0cm 0cm 0cm 0cm},clip,height=0.32\textwidth]{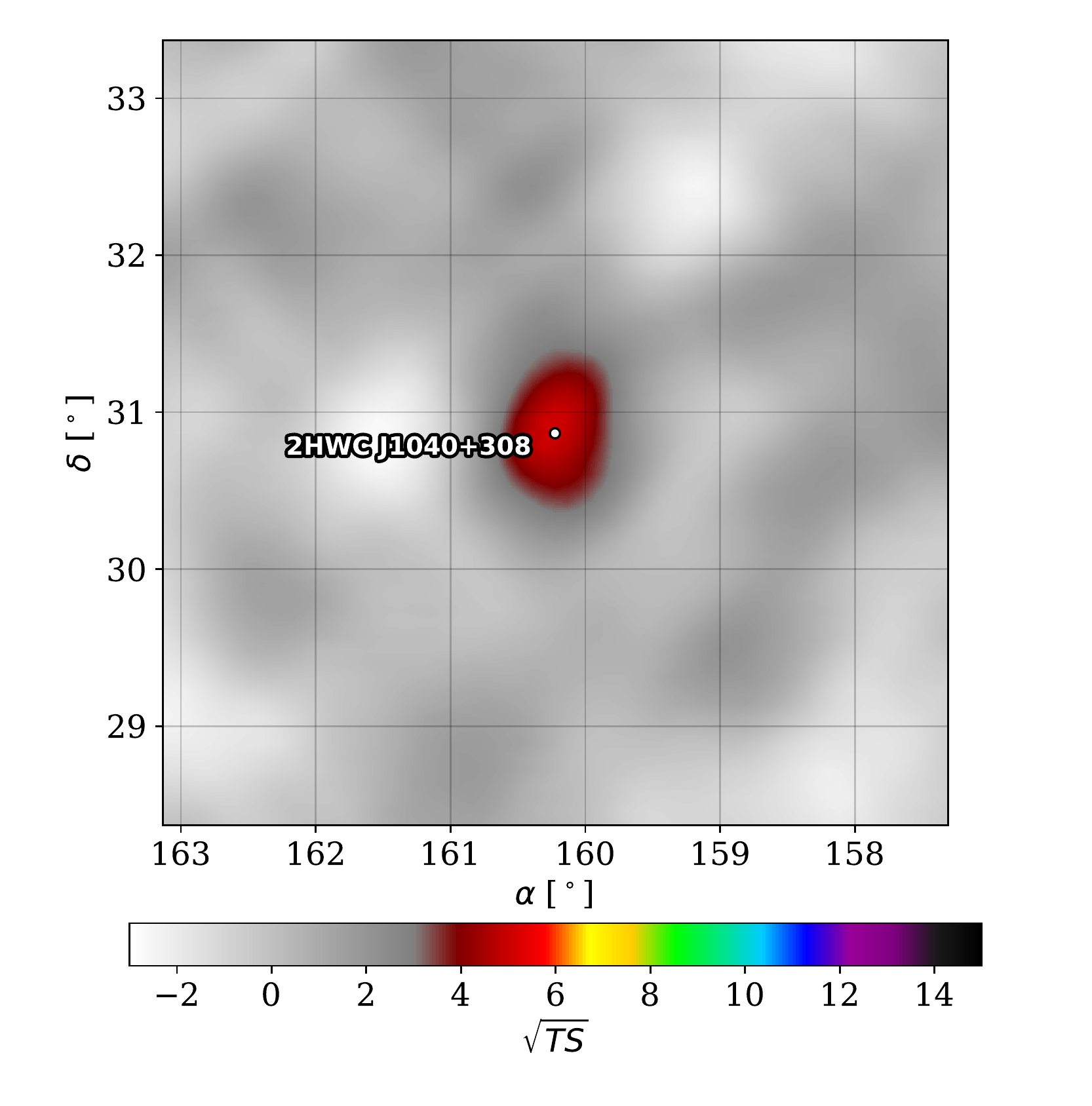}\\
    \includegraphics[trim={0cm 0cm 0cm 0cm},clip,height=0.32\textwidth]{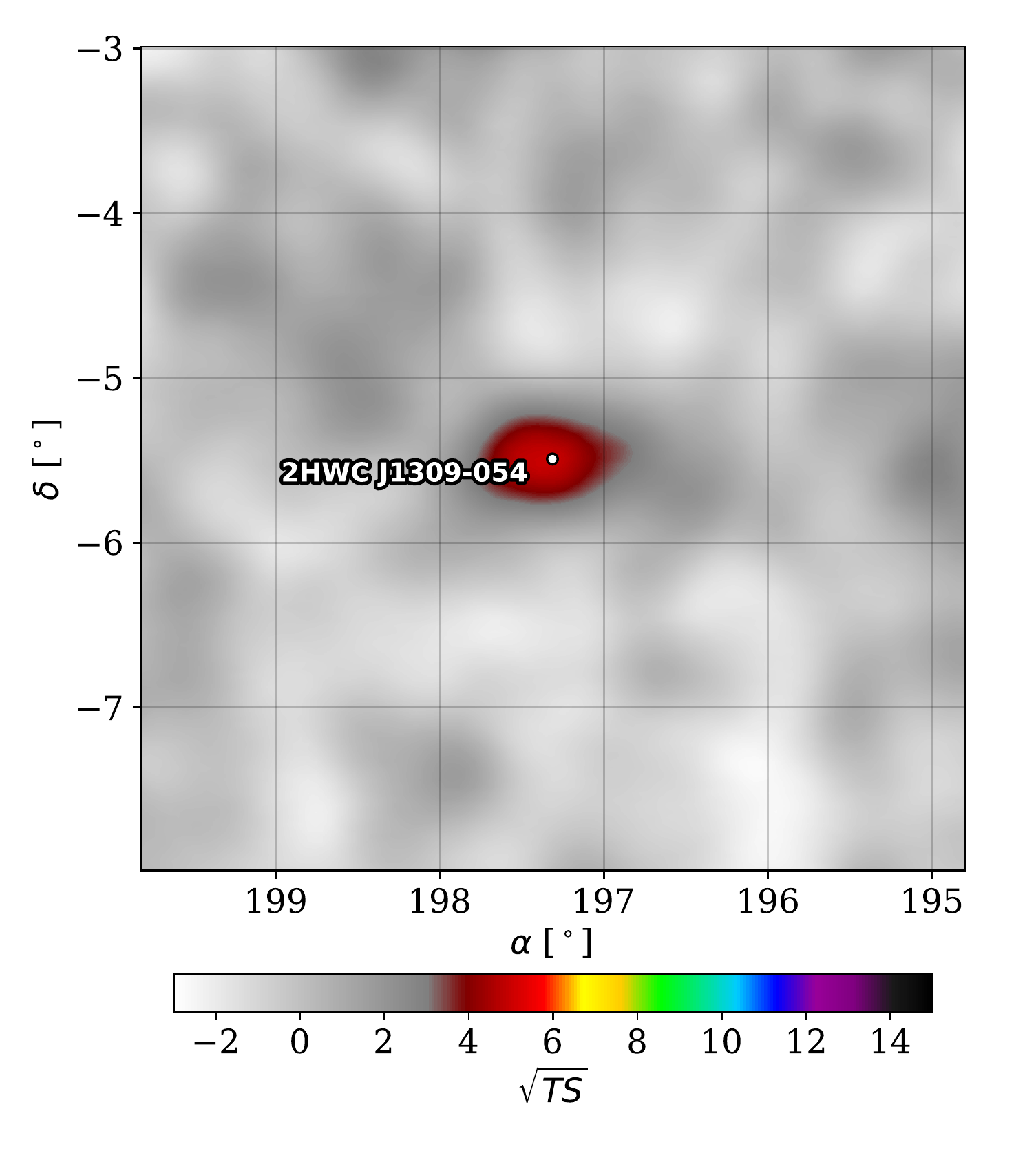}
    \includegraphics[trim={0cm 0cm 0cm 0cm},clip,height=0.32\textwidth]{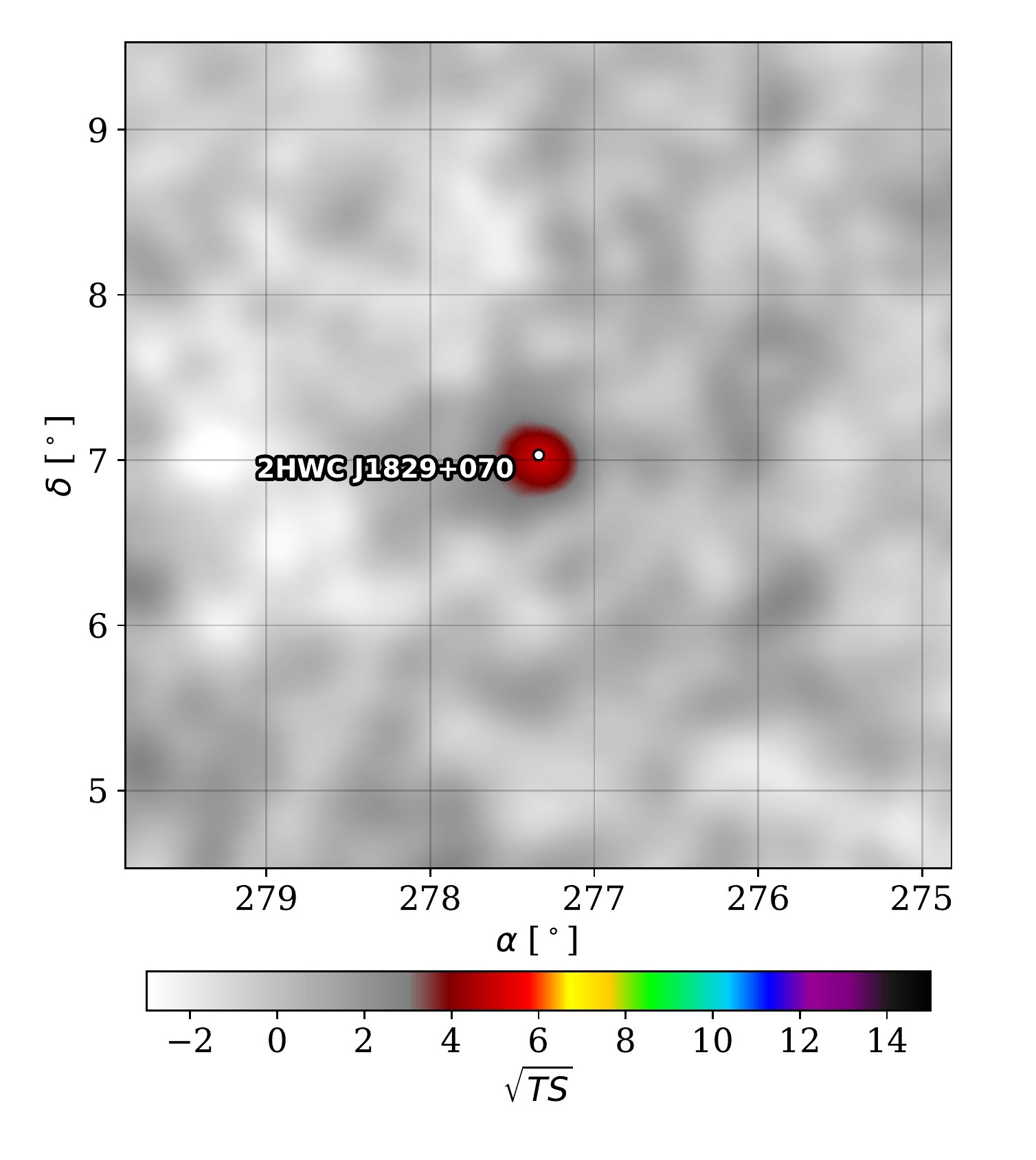}
    \caption{Regions around 2HWC J0819+157, 2HWC J1040+308, 2HWC J1309-054, and 2HWC J1829+070 in equatorial
             coordinates. The TS maps correspond to the search in which these
             sources were found: the extended source hypothesis with a radius
             of 0.5\degree and a spectral index of $-2.0$ for the former two, and 
             the point source hypothesis and a spectral index of $-2.7$ for the latter
             two.}
    \label{fig:extragal}
  \end{figure*}

\begin{figure*}
    \centering
    \includegraphics[trim={0cm 3.2cm 0cm 1cm},clip,width=\textwidth]{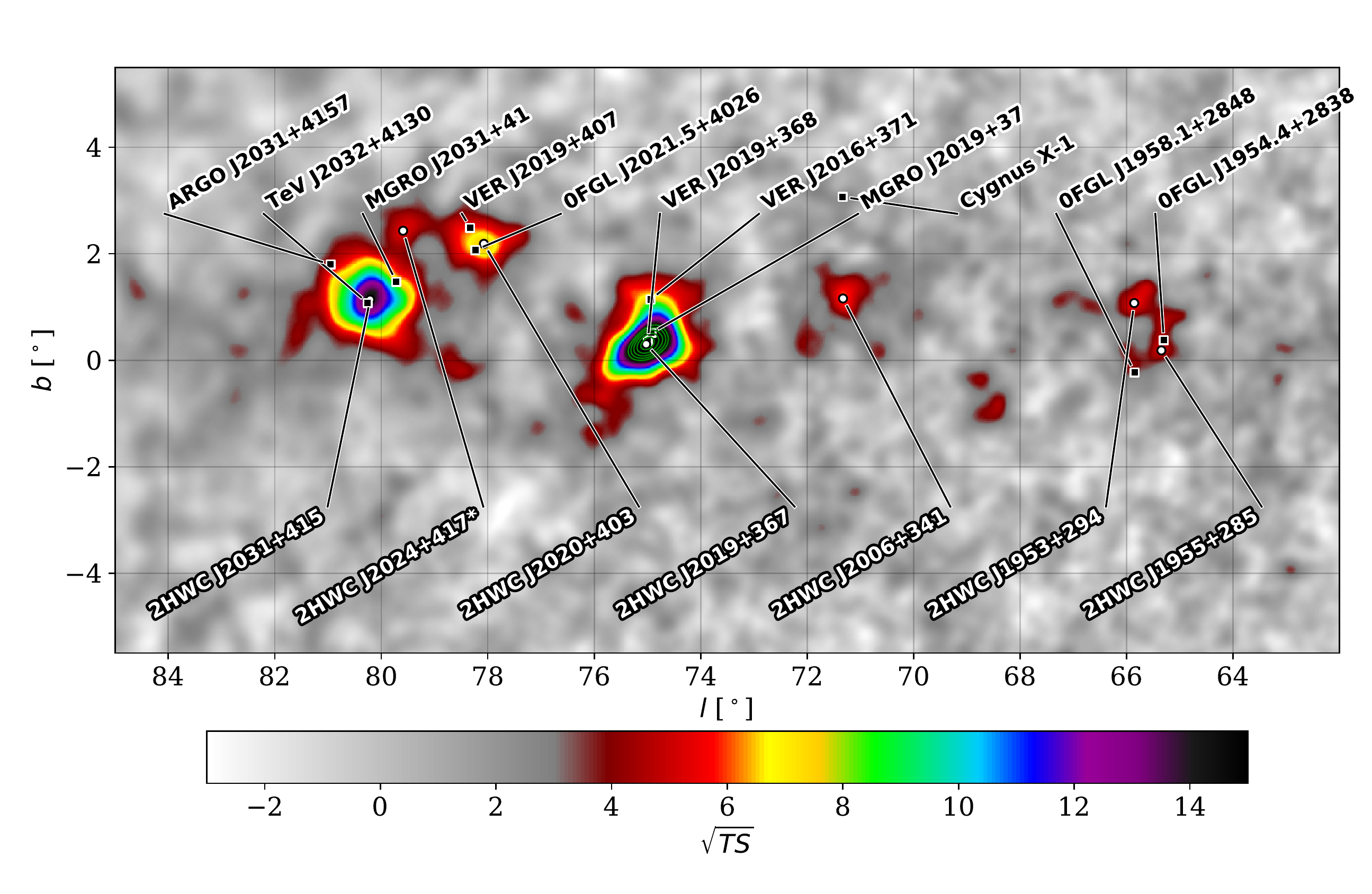}
    \includegraphics[trim={0cm 0cm 0cm 1cm},clip,width=\textwidth]{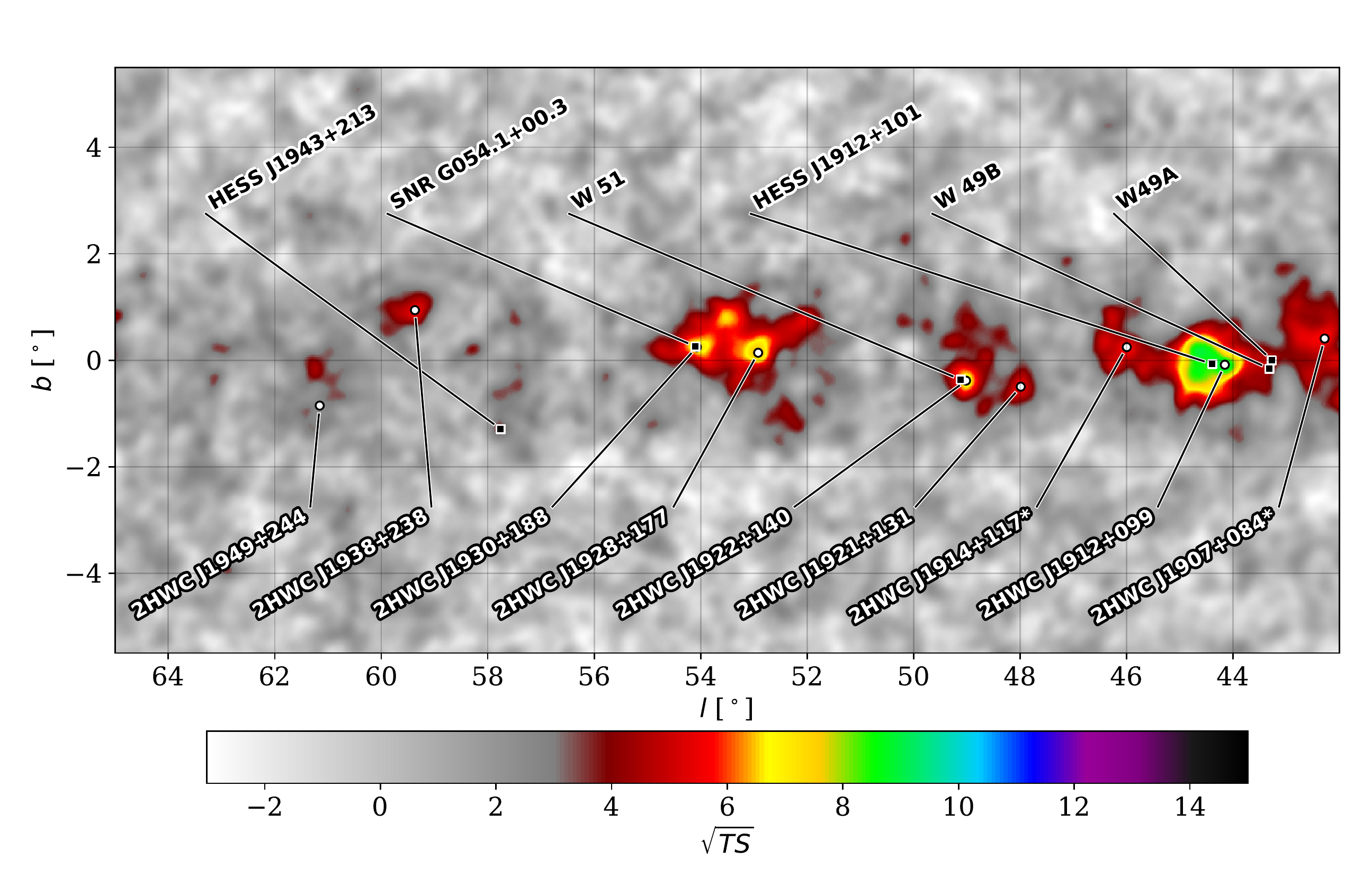}
    \caption{Parts of the inner Galactic Plane region, in Galactic coordinates.
    The TS map corresponds to a point source hypothesis with a spectral index
    of $-2.7$. The green contour lines indicate values of \sts of 15, 16, 17, etc.
    In this figure and the following, the 2HWC sources are represented by white
    circles and labels below the circle; whereas the source listed in TeVCat
    are represented with black squares and labels above the square symbol.}
    \label{fig:gp01}
  \end{figure*}

\begin{figure*}
    \centering
    \includegraphics[trim={0cm 3.2cm 0cm 1cm},clip,width=\textwidth]{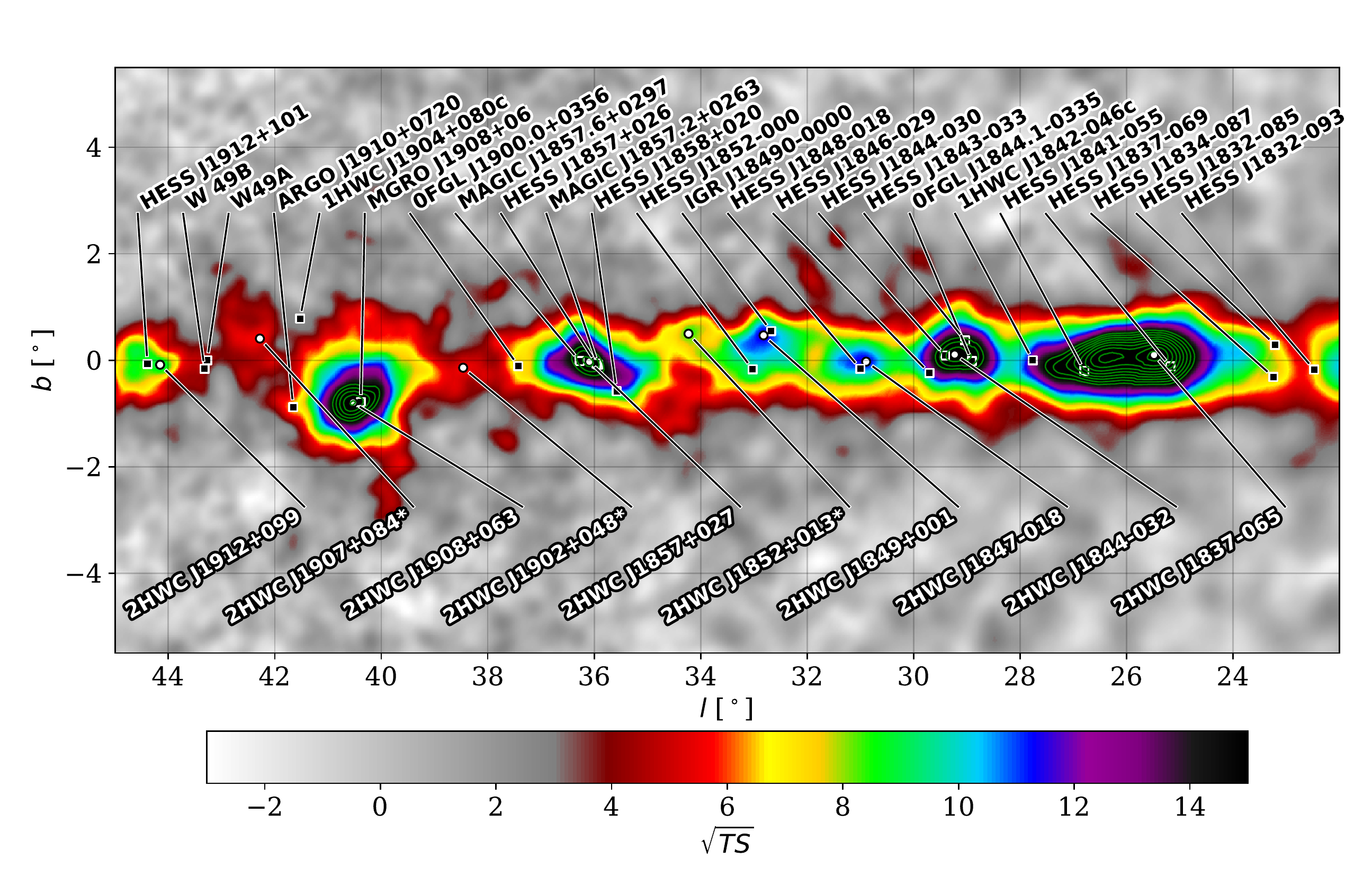}
    \includegraphics[trim={0cm 0cm 0cm 1cm},clip,width=\textwidth]{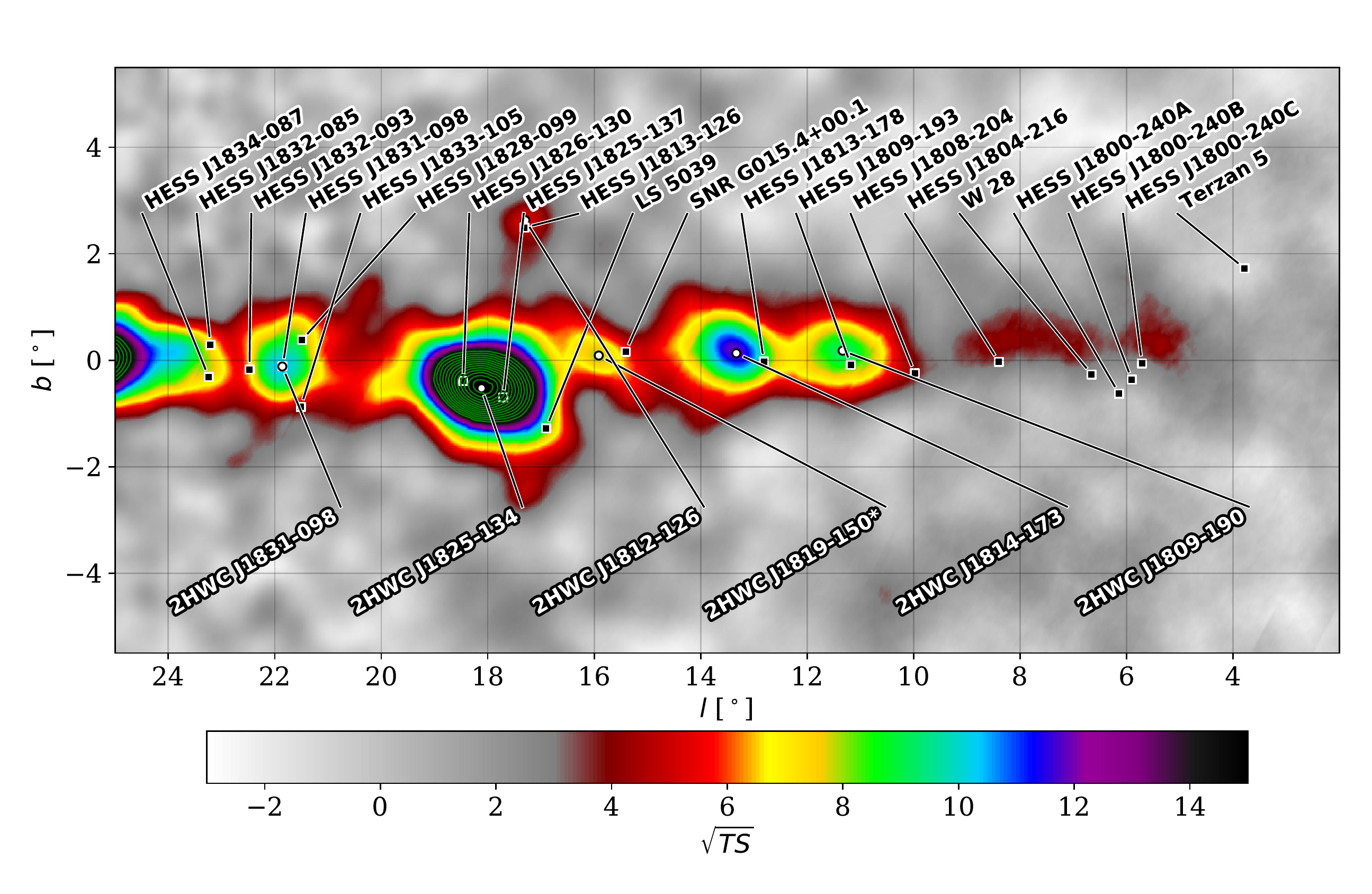}
    \caption{Same as Figure~\ref{fig:gp01}, farther along the Galactic Plane.}
    \label{fig:gp23}
  \end{figure*}

  Figures \ref{fig:crab} to \ref{fig:gp23} show detailed views of smaller regions
  of the sky.
  2HWC sources are represented by white circles and labels
  below the circle.
  The source locations listed in TeVCat
  are also marked, with black squares and labels
  above the square symbol.

  The maps of the regions around the Crab, Markarian~421, and Markarian~501 are shown in
  Figure~\ref{fig:crab}.
  The region of the outer Galactic Plane around Geminga is
  mapped in Figure~\ref{fig:cgg}.
  The left map shows the result of the point source search;
  the right map that of the 2\degree extended search. The increased 
  TS in the extended search supports the case of a significant extent of the two
  TeV sources detected by the HAWC Observatory in this region. 
  Isolated sources found out of the Galactic Plane are shown on Figure~\ref{fig:extragal}.
  Finally, the inner Galactic Plane from the Cygnus region towards the center of the 
  Galaxy is shown in Figures~\ref{fig:gp01} and \ref{fig:gp23}.
  
\subsection{Catalog}\label{sec:results-catalog}
  
  Table~\ref{tab:sources} lists all sources found using the procedure
  described in Section~\ref{sec:method-catalog}, ordered by right ascension.
  The first column lists the HAWC catalog name.
  The second column specifies the search in which the source first appeared
  with a TS above the threshold value of 25.
  PS denotes the point source search, 0.5, 1, and 2\degree the radius of the
  disk in the extended search.
  The corresponding
  TS value is reported in the third column.
  The following columns compile the source positions in equatorial (J2000.0
  epoch) and Galactic coordinates and the one-sigma uncertainty on the position
  of the maximum identified in the respective search.
  The second part of the table, after the vertical line, provides information on
  the nearest TeVCat source: the distance, then the corresponding name if this
  distance is less than 1\degree.

  \begin{deluxetable*}{c c c c c c c c | c c h h h}
    \tabletypesize{\footnotesize}
    \tablecaption{2HWC source list and nearest TeVCat sources. The sources with
      a * symbol correspond to sources that are not separated from their
      neighbor by a large TS gap, as defined in section~\ref{sec:search}.
      \label{tab:sources}}
    \tablehead{
      \colhead{} & \colhead{} & \colhead{} & \colhead{} & \colhead{} & \colhead{} & \colhead{} & \colhead{} & \multicolumn{2}{c}{Nearest TeVCat source}\\
      \cline{9-10}
      \colhead{Name} & \colhead{Search} & \colhead{TS} & \colhead{RA} & \colhead{Dec} & \colhead{l} & \colhead{b} & \colhead{1$\sigma$ stat. unc.} & \colhead{Dist.} & \colhead{Name} & \nocolhead{TeVCat flux} & \nocolhead{TeVCat index} & \nocolhead{TeVCat extent}\\
      \colhead{} & \colhead{} & \colhead{} & \colhead{[$^\circ$]} & \colhead{[$^\circ$]} & \colhead{[$^\circ$]} & \colhead{[$^\circ$]} & \colhead{[$^\circ$]} & \colhead{[$^\circ$]} & \colhead{} & \nocolhead{[CU]} & \nocolhead{} & \nocolhead{[$^\circ$]}
    }
    \startdata
       2HWC~J0534+220 &           PS & 1.1E+4 &  83.63 &  22.02 & 184.55 &  -5.78 &  0.06 &  0.01 &            Crab & 1.000 & -2.50 &   No \\
 2HWC~J0631+169 &           PS &   29.6 &  98.00 &  17.00 & 195.61 &   3.51 &  0.11 &  0.39 &         Geminga & 0.230 &   N/A & 2.60 \\
 2HWC~J0635+180 &           PS &   27.4 &  98.83 &  18.05 & 195.04 &   4.70 &  0.13 &  0.97 &         Geminga & 0.230 &   N/A & 2.60 \\
 2HWC~J0700+143 &  1.0$^\circ$ &     29 & 105.12 &  14.32 & 201.10 &   8.44 &  0.80 &  2.98 &               - &     - &     - &    - \\
 2HWC~J0819+157 &  0.5$^\circ$ &   30.7 & 124.98 &  15.79 & 208.00 &  26.52 &  0.17 &  7.86 &               - &     - &     - &    - \\
 2HWC~J1040+308 &  0.5$^\circ$ &   26.3 & 160.22 &  30.87 & 197.59 &  61.31 &  0.22 &  8.77 &               - &     - &     - &    - \\
 2HWC~J1104+381 &           PS & 1.15E+3 & 166.11 &  38.16 & 179.95 &  65.05 &  0.06 &  0.04 &   Markarian 421 & 0.300 & -2.20 &   No \\
 2HWC~J1309-054 &           PS &   25.3 & 197.31 &  -5.49 & 311.11 &  57.10 &  0.22 &  3.27 &               - &     - &     - &    - \\
 2HWC~J1653+397 &           PS &    556 & 253.48 &  39.79 &  63.64 &  38.85 &  0.07 &  0.03 &   Markarian 501 &   N/A & -2.72 &   No \\
 2HWC~J1809-190 &           PS &   85.5 & 272.46 & -19.04 &  11.33 &   0.18 &  0.17 &  0.31 &  HESS J1809-193 & 0.140 & -2.20 & 0.53 \\
 2HWC~J1812-126 &           PS &   26.8 & 273.21 & -12.64 &  17.29 &   2.63 &  0.19 &  0.14 &  HESS J1813-126 & 0.042 &   N/A & 0.21 \\
 2HWC~J1814-173 &           PS &    141 & 273.52 & -17.31 &  13.33 &   0.13 &  0.18 &  0.54 &  HESS J1813-178 & 0.060 & -2.09 & 0.04 \\
2HWC~J1819-150* &           PS &   62.9 & 274.83 & -15.06 &  15.91 &   0.09 &  0.16 &  0.51 & SNR G015.4+00.1 & 0.018 & -2.10 &   No \\
 2HWC~J1825-134 &           PS &    767 & 276.46 & -13.40 &  18.12 &  -0.53 &  0.09 &  0.39 &  HESS J1826-130 & 0.033 &   N/A & 0.15 \\
 2HWC~J1829+070 &           PS &   25.3 & 277.34 &   7.03 &  36.72 &   8.09 &  0.10 &  8.12 &               - &     - &     - &    - \\
 2HWC~J1831-098 &           PS &    107 & 277.87 &  -9.90 &  21.86 &  -0.12 &  0.17 &  0.01 &  HESS J1831-098 & 0.040 & -2.10 & 0.15 \\
 2HWC~J1837-065 &           PS &    549 & 279.36 &  -6.58 &  25.48 &   0.10 &  0.06 &  0.37 &  HESS J1837-069 & 0.132 & -2.27 & 0.12 \\
 2HWC~J1844-032 &           PS &    309 & 281.07 &  -3.25 &  29.23 &   0.11 &  0.10 &  0.18 &  HESS J1844-030 & 0.010 &   N/A &   No \\
 2HWC~J1847-018 &           PS &    132 & 281.95 &  -1.83 &  30.89 &  -0.03 &  0.11 &  0.17 &  HESS J1848-018 & 0.020 & -2.80 & 0.32 \\
 2HWC~J1849+001 &           PS &    134 & 282.39 &   0.11 &  32.82 &   0.47 &  0.10 &  0.16 & IGR J18490-0000 & 0.015 &   N/A &  Yes \\
2HWC~J1852+013* &           PS &   71.4 & 283.01 &   1.38 &  34.23 &   0.50 &  0.13 &  1.37 &               - &     - &     - &    - \\
 2HWC~J1857+027 &           PS &    303 & 284.33 &   2.80 &  36.09 &  -0.03 &  0.06 &  0.14 &  HESS J1857+026 &   N/A & -2.39 & 0.11 \\
2HWC~J1902+048* &           PS &   31.7 & 285.51 &   4.86 &  38.46 &  -0.14 &  0.18 &  2.03 &               - &     - &     - &    - \\
2HWC~J1907+084* &           PS &   33.1 & 286.79 &   8.50 &  42.28 &   0.41 &  0.27 &  1.15 &               - &     - &     - &    - \\
 2HWC~J1908+063 &           PS &    367 & 287.05 &   6.39 &  40.53 &  -0.80 &  0.06 &  0.14 &   MGRO J1908+06 & 0.170 & -2.10 & 0.34 \\
 2HWC~J1912+099 &           PS &   83.2 & 288.11 &   9.93 &  44.15 &  -0.08 &  0.10 &  0.24 &  HESS J1912+101 & 0.100 & -2.70 & 0.26 \\
2HWC~J1914+117* &           PS &     33 & 288.68 &  11.72 &  46.00 &   0.25 &  0.13 &  1.64 &               - &     - &     - &    - \\
 2HWC~J1921+131 &           PS &   30.1 & 290.30 &  13.13 &  47.99 &  -0.50 &  0.12 &  1.14 &               - &     - &     - &    - \\
 2HWC~J1922+140 &           PS &     49 & 290.70 &  14.09 &  49.01 &  -0.38 &  0.11 &  0.10 &            W 51 & 0.030 &   N/A & 0.12 \\
 2HWC~J1928+177 &           PS &   65.7 & 292.15 &  17.78 &  52.92 &   0.14 &  0.07 &  1.18 &               - &     - &     - &    - \\
 2HWC~J1930+188 &           PS &   51.8 & 292.63 &  18.84 &  54.07 &   0.24 &  0.12 &  0.03 & SNR G054.1+00.3 & 0.025 & -2.39 &   No \\
 2HWC~J1938+238 &           PS &   30.5 & 294.74 &  23.81 &  59.37 &   0.94 &  0.13 &  2.75 &               - &     - &     - &    - \\
 2HWC~J1949+244 &  1.0$^\circ$ &   34.9 & 297.42 &  24.46 &  61.16 &  -0.85 &  0.71 &  3.43 &               - &     - &     - &    - \\
 2HWC~J1953+294 &           PS &   30.1 & 298.26 &  29.48 &  65.86 &   1.07 &  0.24 &  8.44 &               - &     - &     - &    - \\
 2HWC~J1955+285 &           PS &   25.4 & 298.83 &  28.59 &  65.35 &   0.18 &  0.14 &  7.73 &               - &     - &     - &    - \\
 2HWC~J2006+341 &           PS &   36.9 & 301.55 &  34.18 &  71.33 &   1.16 &  0.13 &  3.61 &               - &     - &     - &    - \\
 2HWC~J2019+367 &           PS &    390 & 304.94 &  36.80 &  75.02 &   0.30 &  0.09 &  0.07 &   VER J2019+368 &   N/A & -1.75 & 0.34 \\
 2HWC~J2020+403 &           PS &   59.7 & 305.16 &  40.37 &  78.07 &   2.19 &  0.11 &  0.40 &   VER J2019+407 & 0.037 & -2.37 & 0.23 \\
2HWC~J2024+417* &           PS &   28.4 & 306.04 &  41.76 &  79.59 &   2.43 &  0.20 &  0.97 &   MGRO J2031+41 & 0.390 & -3.22 & 1.80 \\
 2HWC~J2031+415 &           PS &    209 & 307.93 &  41.51 &  80.21 &   1.14 &  0.09 &  0.08 &  TeV J2032+4130 & 0.030 & -2.00 & 0.09 \\

    \enddata
  \end{deluxetable*}

  Table~\ref{tab:fluxes} lists the differential photon flux at 7\,TeV ($F_7$) and the
  spectral index of the power law that fit the source identified in HAWC data
  best.
  For all sources we report the flux estimated with the source model
  corresponding to the search in which the source was found.
  For the sources for which an additional source size hypothesis was defined, as
  detailed in Section~\ref{sec:parameter-estimate}, the second flux measurement
  is also reported.

  \begin{deluxetable*}{c c c D@{\hspace{0.35em}$\pm$\hspace{-0.3em}}D c}
    \tabletypesize{\ssmall} 
    \tablecaption{The 2HWC catalog: Source radius, fitted spectrum, and TeV
      counterpart.
      The flux $F_7$ is the differential flux at 7\,TeV.
      For some sources an additional line indicates another spectral fit
      with a more extended source assumption.
      The uncertainties reported here are statistical only.
      The systematic uncertainties are 0.1\degree for the position, 50\%
      for the flux, and 0.2 for the index.
      \label{tab:fluxes}}
    \tablehead{
      \colhead{Name} & \colhead{Tested radius} & \colhead{Index} & \multicolumn{4}{c}{$F_{7} \times 10^{15}$} & \colhead{TeVCat}\\
      \colhead{} & \colhead{[$^\circ$]} & \colhead{} & \multicolumn{4}{c}{[TeV$^{-1}$cm$^{-2}$s$^{-1}$]} & \colhead{}
    }
    \startdata
      \decimals
       2HWC~J0534+220 &   - & -2.58 $\pm$  0.01 &       184.7 &     2.4 & Crab \\
 2HWC~J0631+169 &   - & -2.57 $\pm$  0.15 &         6.7 &     1.5 & Geminga \\
              " & 2.0 & -2.23 $\pm$  0.08 &        48.7 &     6.9 & Geminga \\
 2HWC~J0635+180 &   - & -2.56 $\pm$  0.16 &         6.5 &     1.5 & Geminga \\
 2HWC~J0700+143 & 1.0 & -2.17 $\pm$  0.16 &        13.8 &     4.2 & - \\
              " & 2.0 & -2.03 $\pm$  0.14 &        23.0 &     7.3 & - \\
 2HWC~J0819+157 & 0.5 & -1.50 $\pm$  0.67 &         1.6 &     3.1 & - \\
 2HWC~J1040+308 & 0.5 & -2.08 $\pm$  0.25 &         6.6 &     3.5 & - \\
 2HWC~J1104+381 &   - & -3.04 $\pm$  0.03 &        70.8 &     2.9 & Markarian 421 \\
 2HWC~J1309-054 &   - & -2.55 $\pm$  0.18 &        12.3 &     3.5 & - \\
 2HWC~J1653+397 &   - & -2.86 $\pm$  0.04 &        56.5 &     2.7 & Markarian 501 \\
 2HWC~J1809-190 &   - & -2.61 $\pm$  0.11 &        80.9 &    15.1 & HESS J1809-193 \\
 2HWC~J1812-126 &   - & -2.84 $\pm$  0.16 &        27.4 &     5.7 & HESS J1813-126 \\
 2HWC~J1814-173 &   - & -2.61 $\pm$  0.09 &        88.4 &    13.0 & HESS J1813-178 \\
              " & 1.0 & -2.55 $\pm$  0.07 &       151.6 &    18.8 & HESS J1813-178 \\
2HWC~J1819-150* &   - & -2.88 $\pm$  0.10 &        59.0 &     7.9 & SNR G015.4+00.1 \\
 2HWC~J1825-134 &   - & -2.58 $\pm$  0.04 &       138.0 &     8.1 & HESS J1826-130 \\
              " & 0.9 & -2.56 $\pm$  0.03 &       249.2 &    11.4 & HESS J1826-130 \\
 2HWC~J1829+070 &   - & -2.69 $\pm$  0.17 &         8.1 &     1.7 & - \\
 2HWC~J1831-098 &   - & -2.80 $\pm$  0.09 &        44.2 &     4.7 & HESS J1831-098 \\
              " & 0.9 & -2.64 $\pm$  0.06 &        95.8 &     8.0 & HESS J1831-098 \\
 2HWC~J1837-065 &   - & -2.90 $\pm$  0.04 &        85.2 &     4.1 & HESS J1837-069 \\
              " & 2.0 & -2.66 $\pm$  0.03 &       341.3 &    11.3 & HESS J1837-069 \\
 2HWC~J1844-032 &   - & -2.64 $\pm$  0.06 &        46.8 &     3.2 & HESS J1844-030 \\
              " & 0.6 & -2.51 $\pm$  0.04 &        92.8 &     5.2 & HESS J1844-030 \\
 2HWC~J1847-018 &   - & -2.95 $\pm$  0.08 &        28.9 &     2.8 & HESS J1848-018 \\
 2HWC~J1849+001 &   - & -2.54 $\pm$  0.10 &        22.8 &     2.9 & IGR J18490-0000 \\
              " & 0.8 & -2.47 $\pm$  0.05 &        60.8 &     4.5 & IGR J18490-0000 \\
2HWC~J1852+013* &   - & -2.90 $\pm$  0.10 &        18.2 &     2.3 & - \\
 2HWC~J1857+027 &   - & -2.93 $\pm$  0.05 &        35.5 &     2.5 & HESS J1857+026 \\
              " & 0.9 & -2.61 $\pm$  0.04 &        97.3 &     4.4 & HESS J1857+026 \\
2HWC~J1902+048* &   - & -3.22 $\pm$  0.16 &         8.3 &     2.4 & - \\
2HWC~J1907+084* &   - & -3.25 $\pm$  0.18 &         7.3 &     2.5 & - \\
 2HWC~J1908+063 &   - & -2.52 $\pm$  0.05 &        34.1 &     2.2 & MGRO J1908+06 \\
              " & 0.8 & -2.33 $\pm$  0.03 &        85.1 &     4.2 & MGRO J1908+06 \\
 2HWC~J1912+099 &   - & -2.93 $\pm$  0.09 &        14.5 &     1.9 & HESS J1912+101 \\
              " & 0.7 & -2.64 $\pm$  0.06 &        36.6 &     3.0 & HESS J1912+101 \\
2HWC~J1914+117* &   - & -2.83 $\pm$  0.15 &         8.5 &     1.6 & - \\
 2HWC~J1921+131 &   - & -2.75 $\pm$  0.15 &         7.9 &     1.5 & - \\
 2HWC~J1922+140 &   - & -2.49 $\pm$  0.15 &         8.7 &     1.8 & W 51 \\
              " & 0.9 & -2.51 $\pm$  0.09 &        26.1 &     3.4 & W 51 \\
 2HWC~J1928+177 &   - & -2.56 $\pm$  0.14 &        10.0 &     1.7 & - \\
 2HWC~J1930+188 &   - & -2.74 $\pm$  0.12 &         9.8 &     1.5 & SNR G054.1+00.3 \\
 2HWC~J1938+238 &   - & -2.96 $\pm$  0.15 &         7.4 &     1.6 & - \\
 2HWC~J1949+244 & 1.0 & -2.38 $\pm$  0.16 &        19.4 &     4.2 & - \\
 2HWC~J1953+294 &   - & -2.78 $\pm$  0.15 &         8.3 &     1.6 & - \\
 2HWC~J1955+285 &   - & -2.40 $\pm$  0.24 &         5.7 &     2.1 & - \\
 2HWC~J2006+341 &   - & -2.64 $\pm$  0.15 &         9.6 &     1.9 & - \\
              " & 0.9 & -2.40 $\pm$  0.11 &        24.5 &     4.2 & - \\
 2HWC~J2019+367 &   - & -2.29 $\pm$  0.06 &        30.2 &     3.1 & VER J2019+368 \\
              " & 0.7 & -2.24 $\pm$  0.04 &        58.2 &     4.6 & VER J2019+368 \\
 2HWC~J2020+403 &   - & -2.95 $\pm$  0.10 &        18.5 &     2.6 & VER J2019+407 \\
2HWC~J2024+417* &   - & -2.74 $\pm$  0.17 &        12.4 &     2.6 & MGRO J2031+41 \\
 2HWC~J2031+415 &   - & -2.57 $\pm$  0.07 &        32.4 &     3.2 & TeV J2032+4130 \\
              " & 0.7 & -2.52 $\pm$  0.05 &        61.6 &     4.4 & TeV J2032+4130 \\

    \enddata
  \end{deluxetable*}%
  
  \begin{figure}
   \begin{center}
     \includegraphics[width=0.45\textwidth]{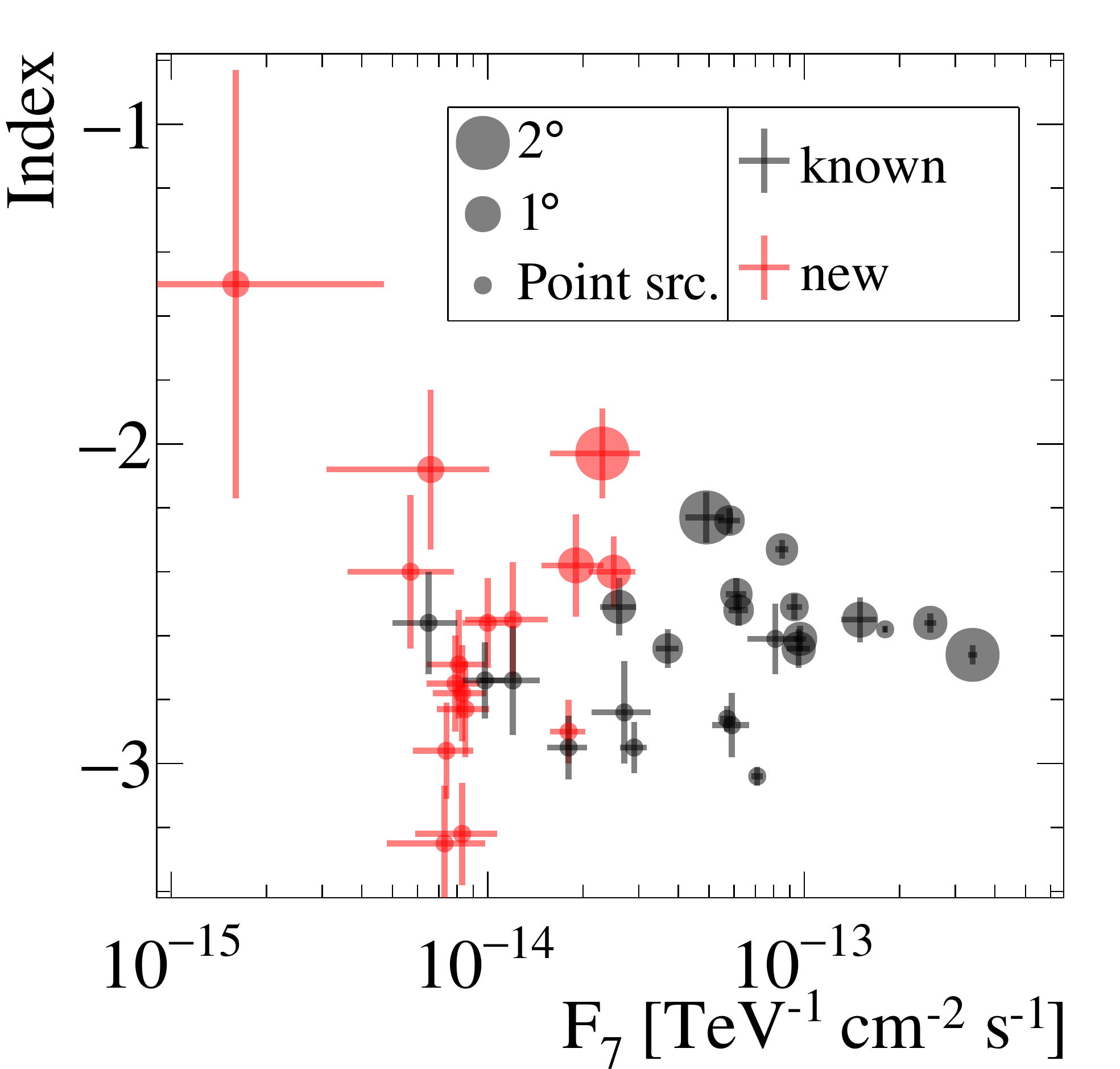}
     \caption{Distribution of the 2HWC sources in flux at 7\,TeV ($F_7$) and
     power-law index. The marker size indicates the source extend to calculate
     the source flux and the color indicates whether these sources have (gray)
     or do not have (red) a counterpart in TeVCat.}
     \label{fig:FluxIndexSize}
   \end{center}
   \end{figure}
   
The results of Table~\ref{tab:fluxes} are illustrated in
Figure~\ref{fig:FluxIndexSize}. For fluxes $F_{7} >
3\times10^{-14}$\dfu all sources have 
previously been detected using other instruments, but below this value the
fraction of newly detected sources dominates the sample.  We note
here that, when taking into account the full extent of each source, the Crab
Nebula is only the third brightest source in the sky at 7\,TeV. The
brightest sources are 2HWC J1837-065 and 2HWC J1825-134.

In Figure~\ref{fig:FluxIndexSize} there is a region, around $F_7 =
0.8\times10^{-15}$\dfu and power law index
$<$$-2.7$, where new catalog sources cluster. These sources do not have significant flux beyond the PSF of HAWC and should
therefore provide interesting targets for follow-up with IACTs.

\section{Discussion}
  \label{sec:discussion}
  In this section we briefly discuss each source and its possible associations,
ordered by right ascension.
Of particular interest are the sources detected with
previous and current TeV instruments, including the 1HWC sources observed in the
inner Galaxy with a partial configuration of HAWC
\citep{hawc111paper} and sources listed in TeVCat.
GeV counterparts are also searched in the \LAT catalogs:
the standard 2FGL and 3FGL catalogs \citep{0067-0049-188-2-405,0067-0049-199-2-31,2015ApJS..218...23A},
the high energy 1FHL and 2FHL catalogs \citep{0067-0049-209-2-34,0067-0049-222-1-5}, the
second pulsar catalogs \citep{2013ApJS..208...17A}, and the
SNR catalog \citep{2016ApJS..224....8A}.
The ATNF pulsar catalog \citep{2005AJ....129.1993M} is used to look for nearby
pulsars. When available, the pulsars spindown power $\dot{E}$, distance $d$, and
age $\tau$ are reported, as obtained from the ATNF catalog unless mentioned otherwise.
Associations are typically search for within 0.5\degree of the position measured
by HAWC.

  \subsection{2HWC J0534+220 -- Crab}
    2HWC J0534+220 is the source with the largest significance in this
    catalog, with $\mathrm{TS}=1.1 \times 10^4$. It corresponds to the Crab PWN,
    which is the first TeV source detected, in 1989 \citep{whipplecrabdiscovery},
    and which is since commonly used as a calibration source for TeV instruments.
    The associated pulsar is young and has a high spindown power
    ($\dot{E}=4.5\times 10^{38}$\ergps, $d=2.0$\,kpc, $\tau=1.26$\,kyr).
    In the GeV regime, the emission is dominated by the pulsed emission originating from the pulsar.
    Although the pulsed emission has been observed up to 1.5\,TeV \citep{2016A&A...585A.133A},
    most of the TeV emission is due to inverse Compton scattering in the surrounding PWN
    \citep{1996MNRAS.278..525A}.
    
    The spectrum measured here matches previously published results.
    A more complete analysis of the Crab Nebula observation by HAWC will be
    presented in a separate publication \citep{crabpaper}.

  \subsection{2HWC J0631+169 and 2HWC J0635+180 -- Geminga}
    2HWC J0631+169 and 2HWC J0635+180 are both found in
    the point source search, each above the TS threshold value of 25.
    The corresponding TS maximum in the 2\degree extended search is 126.
    They appear to be associated with Geminga, a known GeV \citep{0004-637X-720-1-272}
    gamma-ray pulsar.
    Prior to HAWC, Milagro was the only TeV instrument to have detected it.
    Milagro reported an extended source of full width at half maximum around 2.6\degree and a hard spectrum \citep{Abdo:2009ku}.
    The large extent of the source makes it difficult for IACTs to observe it.
    To date none have reported a detection of Geminga (see e.g. \citet{Ahnen:2016ujd}).
    
    Compared to other TeV PWNe, the associated pulsar PSR J0633+1746 is
    relatively old (342\,kyr), nearby ($250^{+120}_{-62}$\,pc) and has a
    low spindown power $(3.2\times 10^{34}$\ergps).
    Geminga (together with PSR B0656+14) has been proposed as the dominant source
    of the local population of TeV electrons and positrons, and thus a possible
    explanation for the PAMELA positron excess \citep{1995A&A...294L..41A,Yuksel:2008rf}.

    When fitted with a uniform disk source model, the extent observed in HAWC is around 2\degree in radius, and the measured
    spectral index is relatively hard at $-2.2$.
    The measured spectrum depends on the assumed morphology.
    A detailed study of Geminga and 2HWC J0700+143 (see next section) by
    HAWC will be presented in a dedicated publication \citep{gemingapaper}.

  \subsection{2HWC J0700+143}
    2HWC J0700+143 is a new TeV source discovered in the 1\degree extended
    search, with a TS of 29.
    The corresponding TS maximum in the 2\degree extended search is 51.
    It is likely associated with the B0656+14 pulsar, which has similar
    characteristics to the Geminga pulsar: old (111\,kyr), nearby
    ($288^{+33}_{-27}$\,pc) and low spindown power ($3.8 \times 10^{34}$\ergps) \citep{Brisken:2003}.
    The associated supernova is believed to be the origin of the Monogem Ring.
    As for Geminga, PSR B0656+14 has been proposed as a significant contributor
    to the local lepton populations.

    The measured extent of this source is around 2\degree, with a hard spectral
    index of about $-2$.

  \subsection{2HWC J0819+157}
    This source is found in the 0.5\degree
    radius extended search, with a TS value of 30.7.
    The coordinates correspond to a location out of the Galactic Plane ($b=26.52$\degree).
    The fitted index ($-1.50$) is much harder than the fitted index of any
    other source.
    The nearest potentially high energy source is the AGN 
    2MASS J08203478+1531114, 0.3 away.
    However, its distance ($z=0.14$) seems incompatible with the observed extent
    and hard spectrum.
    
  \subsection{2HWC J1040+308}
    Similar to 2HWC J0819+157, this source is found in the 0.5\degree
    radius extended search, with a TS value of 26.3.
    No obvious associations are found in the catalogs.
    The coordinates correspond to a location out of the Galactic Plane
    ($b=61.31^\circ$), which seems in tension with the source extent.

  \subsection{2HWC J1104+381 and 2HWC J1653+397 -- Markarian 421 and Markarian 501}
    Markarian (Mrk) 421 and Mrk 501 are two of the closest and brightest
    \extragal sources in the TeV as well as the X-ray band.
    The locations of these two sources (2HWC J1104+381 for Mrk 421 and 
    2HWC J1653+397 for Mrk 501) are the only ones in this catalog that have 
    confirmed \extragal associations.
    
    At a distance of $z\approx 0.031$ \citep{1991rc3..book.....D,2MASSz}, Mrk 421 is
    a BL Lac type blazar that was the first \extragal object discovered 
    at very high energies \citep{1992Natur.358..477P} and has been 
    extensively studied in
    both the spectral and time domains.
    
    Mrk 501 is also a BL Lac type blazar, at a distance of
    $z=0.033$ \citep{1991rc3..book.....D,2MASSz}.
    This object was the second blazar to be detected at very high energies
    \citep{1996ApJ...456L..83Q} and is on average the second brightest \extragal
    object emitting in the TeV band.

    The fluxes of both objects are known to exhibit strong variability on time 
    scales down to hours or even minutes; see for example \citet{1996Natur.383..319G} 
    for Mrk 421 or \citet{Albert2007Mrk501} for Mrk 501.
    A first look at week-long VHE flares and the time dependence of their 
    emission observed with the partial HAWC detector is reported 
    in \citet{blazars-icrc2015}.
    Both higher and lower yearly average fluxes for Mrk 421 than the one 
    listed in Table~\ref{tab:fluxes} have been reported
    in the past \citep{2014APh....54....1A}.
    A detailed characterization of the VHE variability of Mrk  421 and Mrk 501 
and a discussion of their spectral features beyond a power law 
    fit will be the topic of a forthcoming HAWC publication, based on the same 
data discussed here but resolved into daily time intervals.

\subsection{2HWC J1309-054}
    This source is found in the point search with a TS value of 25.3.
    No obvious associations are found in the catalogs.
    The coordinates correspond to a location out of the Galactic Plane
    ($b=57.1^\circ$).

\subsection{2HWC J1809-190}
  2HWC J1809-190 may be associated with HESS J1809-193 (centered
  ${\sim} 0.3^\circ$ away) \citep{2007A&A...472..489A}.
  H.E.S.S. observed it as an extended source modeled with an ellipse of major 
  and minor axis 0.53\degree and 0.25\degree respectively. 
  Suzaku observations confirmed hard extended X-ray emission previously detected
  by ASCA and suggested a possible PWN origin \citep{2010PASJ...62..179A}. 
  However, subsequent radio observations with the Expanded Very Large Array at
  1.4\,GHz suggested that the gamma-ray emission could instead originate from 
  a system of molecular clouds on the edge of the SNR G11.0-0.0 shock front
  \citep{2016A&A...587A..71C} and the gamma source is still considered
  unidentified.

\subsection{2HWC J1812-126}
  2HWC J1812-126 may be associated with the TeV source HESS J1813-126
  (distance of ${\sim} 0.1^\circ$).
  HESS J1813-126 was recently discovered by the H.E.S.S. experiment
  \citep{ICRC-HESS-Legacy-Survey} and is still unidentified.
  The intermediate age pulsar PSR J1813-1246, which has been also detected by \LAT,
 seems coincident with the position of
  the H.E.S.S. source and has a spindown luminosity
  $\dot{E} = 6.2 \times 10^{36}$\ergps and a characteristic age of 43\,kyr.

\subsection{2HWC J1814-173}
  2HWC J1814-173 is close by and possibly associated with the TeV source HESS J1813-178
  (distance of ${\sim} 0.5^\circ$), which was detected during the first H.E.S.S. Galactic Plane survey
  \citep{2005Sci...307.1938A,2006ApJ...636..777A}. 
  HESS J1813-178 is a candidate PWN, powered by the highly energetic young pulsar PSR J1813-1749 located close
  to the center of supernova remnant G12.82-0.02
  \citep{2009ApJ...700L.158G}.
  PSR J1813-1749 has a spindown luminosity of
  $\dot{E} = 6.8 \times 10^{37}$\ergps, a characteristic
  age of 3.3--7.5\,kyr \citep{2009ApJ...700L.158G}, and an estimated
  distance of 4.8\,kpc \citep{2012ApJ...753L..14H}.
  Closer to the measured HAWC location is SNR G013.5+00.2 (0.2\degree away),
  though it has not been detected in gamma rays by H.E.S.S. or \LAT.

\subsection{2HWC J1819-150*}
  2HWC J1819-150* is 0.5\degree away from the nearest source listed in TeVCat,
  SNR G015.4+00.1 (HESS J1818-154). This source is reported by H.E.S.S. as a point
  source, which given the distance to the HAWC location makes the
  association uncertain.
  Closer to the measured HAWC location is SNR G015.9+00.2 (0.1\degree away),
  though it has not been detected in gamma rays by H.E.S.S. or \LAT.
  There are also 5 ATNF pulsars within 0.5\degree from 2HWC J1819-150*:
  PSR J1819-1458 (${\sim} 0.1^\circ$, $\dot{E}=2.9\times 10^{32}$\ergps, $d=3.3$\,kpc, $\tau=117$\,kyr),
  PSR J1819-1510 (${\sim} 0.2^\circ$, $\dot{E}=2.7\times 10^{31}$\ergps, $d=4.1$\,kpc, $\tau=457$\,Myr),
  PSR J1818-1448 (${\sim} 0.3^\circ$, $\dot{E}=1.1\times 10^{34}$\ergps, $d=5.0$\,kpc, $\tau=725$\,kyr),
  PSR J1818-1519 (${\sim} 0.4^\circ$, $\dot{E}=2.0\times 10^{32}$\ergps, $d=5.4$\,kpc, $\tau=3.6$\,Myr), and
  PSR J1817-1511 (${\sim} 0.4^\circ$, $\dot{E}=5.0\times 10^{33}$\ergps, $d=7.3$\,kpc, $\tau=2.5$\,Myr).

\subsection{2HWC J1825-134}
  2HWC J1825-134 was previously detected by HAWC as 1HWC J1825-133. 
  2HWC J1825-134 is located between two previously reported TeV sources,
  HESS J1825-137 and HESS J1826-130, at about 0.4\degree from both.
  HESS J1826-130 was recently announced by the H.E.S.S. experiment
  \citep{ICRC-HESS-Legacy-Survey}
  and is still unidentified.
  HESS J1825-137 was detected by H.E.S.S. \citep{2005Sci...307.1938A} and was
  identified as a PWN \citep[e.g.][]{2005A&A...442L..25A}.
  It is connected to the
  energetic pulsar PSR J1826-1334 (0.2\degree away from 2HWC J1825-134,
  $\dot{E}=2.8\times 10^{36}$\ergps, $d=3.6$\,kpc, $\tau=21$\,kyr).
  It is generally considered the prototype of offset PWNe.
  HESS J1825-137 shows an energy dependent morphology
  at VHE gamma rays towards the south of the pulsar PSR J1826-1334
  \citep{2006A&A...460..365A}.
  The PWN identification was later confirmed by X-ray observations
  \citep{2008ApJ...675..683P,2009PASJ...61S.189U}
  showing a clear detection of an extended PWN.
  The energy dependent morphology studies of HESS J1825-137 continued in the
  \LAT era \citep{2011ApJ...738...42G,2013ApJ...773...77A}, strengthening
  the key role of this source in understanding the physics of PWNe.
  The extension of the TeV spectrum at higher energies by HAWC is in
  line with this scenario.
  With more HAWC data, future analysis including multiple source fit will
  help disentangle the different components contributing to 2HWC J1825-134.
  
  We note that in the present map, the TeV binary LS~5039 is 1.4\degree away
  from 2HWC J1825-134 and is included in its TS halo in the maps presented here.
  Dedicated studies are being developed to separate emission from LS~5039 from
  2HWC J1825-134.

\subsection{2HWC J1829-070}
  This source is found in the point search with a TS value of 25.3.
  It is located slightly off the Galactic Plane at $b=8.09^\circ$, and no
  associations are found in the catalogs within a 0.5\degree radius.
  
\subsection{2HWC J1831-098}
  2HWC J1831-098 may be associated with the TeV source HESS J1831-098
  (distance of 0.01\degree).
  HESS J1831-098 was detected by the H.E.S.S. experiment in 2011
  \citep{2011ICRC....7..244S}, and is a candidate PWN powered by the nearby
  67\,ms pulsar PSR J1831-0952
  ($\dot{E}=1.1\times 10^{36}$\ergps, $d=3.7$\,kpc, $\tau=128$\,kyr).
  The differential flux at 7\,TeV measured by HAWC is two to five times larger
  than the one reported by H.E.S.S., depending on the source size used in the 
  spectrum fit. The indices measured by HAWC are also softer than the value reported by H.E.S.S., $-2.1 \pm 0.1$.

\subsection{2HWC J1837-065}
  2HWC J1837-065 is the principal maximum of an elongated region containing
  multiple known extended sources which are not resolved in the present analysis.
  2HWC J1837-065 may be associated with the close by TeV source HESS J1837-069
  (distance of ${\sim} 0.4^\circ$).
  HESS J1837-069 can be considered a candidate
  PWN \citep{2006ApJ...636..777A,2013arXiv1306.6833T}.
  This elongated HAWC region also covers the location of the unidentified H.E.S.S.
  source HESS J1841-055, which is a very complex TeV gamma-ray source with many
  potential counterparts, including two SNRs (Kes 73, G26.6-0.1), three
  high spindown pulsars:
  PSR J1841-0524 ($\dot{E}=1\times 10^{35}$\ergps, $d=4.1$\,kpc, $\tau=30$\,kyr),
  PSR J1838-0549 ($\dot{E}=1\times 10^{35}$\ergps, $d=4.0$\,kpc, $\tau=112$\,kyr), and
  PSR J1837-0604 ($\dot{E}=2\times 10^{33}$\ergps, $d=4.8$\,kpc, $\tau=34$\,kyr),
  and an X-ray binary (AX J1841.0-0536).
  ARGO-YBJ  also detected emission from this region, ARGO J1839-0627
  \citep{2013arXiv1303.1258T}.
  This HAWC region will be studied further in a dedicated analysis.

\subsection{2HWC J1844-032}
  2HWC J1844-032 was previously reported by HAWC as 1HWC J1844-031c.
  It has two positionally compatible TeV gamma-ray sources: HESS J1844-030
  (${\sim} 0.2^\circ$ distance) and HESS J1843-033 (${\sim} 0.3^\circ$ distance).
  The TeV detected, well studied, PWN Kes 75
  \citep{2008ICRC....2..823D} is slightly offset from the HAWC source
  (0.6\degree away).
  HESS J1844-030 was recently announced by the H.E.S.S. experiment
  \citep{ICRC-HESS-Legacy-Survey}
  and is still unidentified.
  The following sources are possible associations: G29.4+0.1, AX J1844.6-0305,
  and PMN J1844-0306; SNR or PWN scenarios are considered reasonable.
  AX J1844.6-0305 was discovered by \citet{2000ApJ...542L..49V} and appears in
  the ASCA GIS data as a bright source and is not yet identified.
  PMN J1844-0306 is a complex radio/IR region as described by
  \citet{2000ApJ...542L..49V}. 
  
  The other nearby TeV known source, HESS J1843-033 \citep{2008ICRC....2..579H},
  is a large source with several possible counterparts.
  A possible X-ray counterpart is
  AX J1843.8-0352 (G28.60.1), which is an SNR with a peculiar morphology.
  Chandra \citep{2003ApJ...588..338U} discovered a new source within
  AX J1843.8-0352, CXO J184357-035441, which exhibits a thin thermal spectrum and
  a jetlike tail.
  Other possibilities could be AX J1845.0-0258, which has been considered
  as an anomalous X-ray pulsar (AXP),
  or SNR G28.8+1.5, whose outer shells may
  interact with some undiscovered molecular clouds.
  Further multiwavelength observations are crucial to identify the origin of
  the VHE emission. 

\subsection{2HWC J1847-018}
  2HWC J1847-018 was previously detected by HAWC as 1HWC J1849-017c. 
  It may be associated with 
  the unidentified TeV gamma-ray source 
  HESS J1848-018 (${\sim} 0.2^\circ$
  distance). HESS J1848-018 was discovered by the H.E.S.S. experiment in the
  extended Galactic Plane Survey. It is located in the direction of,
  but slightly offset from, the star-forming region W 43 and hence a possible
  association with it was suggested in \citet{2008AIPC.1085..372C}.
  However the association with the star-forming region has not been further
  confirmed and this source is now considered to be a candidate PWN following
  recent observations by \LAT \citep{2013ApJ...773...77A}.
  Further multiwavelength studies are needed to properly identify the source.

\subsection{2HWC J1849+001}
  2HWC J1849+001 may be associated with the extended TeV source
  HESS J1849-000 (${\sim} 0.2^\circ$ distance)
  \citep{2008AIPC.1085..312T}, which is coincident with the INTEGRAL source
  IGR J18490-0000.
  Further X-ray observations by XMM-Newton and RXTE revealed that
  IGR J18490-0000 is a Pulsar/PWN system ,
  where a young and very energetic pulsar ($\dot{E}=9.8\times 10^{36}$\ergps,
  $\tau=43$\,kyr, distance unknown) is powering
  the system and a compact PWN is detected in the X-ray observations
  \citep{2011ApJ...729L..16G}. 

\subsection{2HWC J1852+013*}
  2HWC J1852+013* is a new TeV detection by HAWC.
  There is no known gamma-ray sources close to this location; the nearest is
  the GeV source 3FGL J1852.8+0158, located 0.6\degree from the
  central position of 2HWC J1852+013*.
  Given the source location, there may be a significant
  contribution of the Galactic diffuse emission to this source.

  Multiwavelength catalog searches reveal several pulsars, several X-ray
  sources and HII regions in the vicinity of
  2HWC J1852+013*.
  Chandra observations exist of a star cluster and infrared dark cloud
  IRDC G34.4+0.23 and NaSt1 (WR 122), a Wolf-Rayet binary. 

  The following pulsars are located close by:
  PSR J1851+0118  (${\sim} 0.1^\circ$, $\dot{E}=7.2\times 10^{33}$\ergps, $d=5.6\,\textrm{kpc}$, $\tau=105$\,kyr) and
  PSR J1850+0124  (${\sim} 0.5^\circ$, $\dot{E}=9.5\times 10^{33}$\ergps, $d=3.4\,\textrm{kpc}$, $\tau=5.2$\,Gyr).

\subsection{2HWC J1857+027}
  2HWC J1857+027 has been previously reported by HAWC as 1HWC J1857+023.
  It may be associated with the close by TeV source HESS J1857+026
  (${\sim} 0.1^\circ$ away) 
  \citep{2008A&A...477..353A}, which was considered a PWN candidate
  \citep[e.g.][]{2011ICRC....6..202T}.
  Recent MAGIC observations revealed that the VHE emission above
  1\,TeV can be spatially separated into two sources: MAGIC J1857.2+0263 and
  MAGIC J1857.6+0297 \citep{2014A&A...571A..96M}.
  They also confirmed the PWN nature of the first source and a molecular cloud
  association was suggested for the second source.
  These two MAGIC sources are too close to be distinguishable in the HAWC
  analysis reported here; but they should be resolved in future analysis
  including simultaneous fit of multiple sources.

\subsection{2HWC J1902+048*}
  2HWC J1902+048* has been tagged by the search algorithm in a region that does
  not have a TeV counterpart.
  However, it appears to be in a confused region, possibly with a large
  contribution of the Galactic diffuse emission, and
  will be better disentangled in future analysis with more data.
  Long Swift observations with a total of 23\,ks have been performed in the
  region of 2HWC J1902+048*, due to gamma-ray burst GRB140610.
  There is no possible counterpart in the 3FGL catalog of \LAT, however
  there are 2 sources from the previous catalogs within 0.5\degree:
  1FGL J1902.3+0503c (0.2\degree away) and 2FGL J1901.1+0427 (0.5\degree away).
  Catalog searches reveal several pulsars,
  several X-ray sources and HII regions in the vicinity of
  2HWC J1902+048*.

  The three closest pulsars in the ATNF catalog are:
  PSR J1901+0459 (${\sim} 0.3^\circ$, $d=12.3$\,kpc),
  PSR J1901+0435 (${\sim} 0.3^\circ$, $\dot{E}=1.0\times 10^{33}$\ergps, $d=10.3$\,kpc, $\tau=1.3$\,Myr), and
  PSR J1901+0510 (${\sim} 0.3^\circ$, $\dot{E}=5.3\times 10^{33}$\ergps, $d=5.9$\,kpc, $\tau=313$\,kyr).
  These pulsars could be powering a PWN which is still undetected due to the
  lack of multiwavelength observations.

\subsection{2HWC J1907+084*}
  2HWC J1907+084* is a new TeV detection by HAWC.
  Given the source location and TS value (33.1), there may be a large
  contribution of the Galactic diffuse emission to this source.
  Multiwavelength catalog searches reveal several pulsars, several X-ray
  sources, HII regions, and a molecular cloud system coincident with or in the vicinity
  of 2HWC J1907+084*. 
  The nearest \LAT source is 3FGL J1904.9+0818, located 0.6\degree away from the
  central position of 2HWC J1907+084*.

  The nearest pulsar from the ATNF catalog is PSR J1908+0839
  (${\sim} 0.3^\circ$ away, $\dot{E}=1.5\times 10^{34}$\ergps,
  $d=8.3$\,kpc, $\tau=1.2$\,Myr).

\subsection{2HWC J1908+063 -- MGRO J1908+06}

  2HWC J1908+063 is associated with the PWN MGRO J1908+06, first discovered
  by the Milagro experiment \citep{2007ApJ...664L..91A} and latter observed
  by \HESS \citep{2009A&A...499..723A},
  ARGO-YBJ \citep{2012ApJ...760..110B},
  VERITAS \citep{2014ApJ...787..166A}, and
  previously by HAWC and reported as 1HWC J1907+062c.
  This source was considered unidentified until the advent of \LAT which shed
  light on the nature of MGRO J1908+06 and strengthened the PWN scenario to
  explain its VHE gamma-ray emission
  \citep{2010ApJ...711...64A, 2013ApJ...773...77A}.
  The spectrum measured in this work (see Table~\ref{tab:fluxes}) under the extended
  hypothesis is
  consistent with the spectra obtained by H.E.S.S., VERITAS, and MILAGRO, and
  lower than the ARGO-YBJ results.

\subsection{2HWC J1912+099}
  2HWC J1912+099 may be associated with the TeV source HESS J1912+101
  (${\sim} 0.2^\circ$ distance), which
  was initially proposed to be a PWN connected to the high spindown luminosity
  pulsar PSR J1913+1011
  ($\dot{E}=2.9\times 10^{36}$\ergps, $d=4.6$\,kpc, $\tau=169$\,kyr)
  \citep{2008A&A...484..435A}. 
  ARGO-YBJ also detected emission from this region, ARGO J1912+1026
  \citep{2013ApJ...779...27B}.
  The spectral index they report is consistent with the one by
  H.E.S.S., but the flux above 1\,TeV is much higher than the value reported by
  H.E.S.S.: in this energy band, the flux of
  the H.E.S.S. source corresponds to ${\sim}$9\% of the Crab Nebula flux,
  while the ARGO-YBJ source flux corresponds to ${\sim}$23\% of the Crab flux.
  This discrepancy occurred for other ARGO-YBJ sources and has been discussed
  in literature \citep{2013ApJ...767...99B}.
  The flux measured with HAWC using the extended source model is in agreement
  with the \HESS measurement.
  Due to the lack of multiwavelength confirmation of the PWN scenario, and
  based on the detection of a shell like morphology seen with increased
  observation time by H.E.S.S., \citet{2015arXiv150903872P} reclassified
  HESS J1912+101 as an SNR candidate.

\subsection{2HWC J1914+117*}
  2HWC J1914+117* is a new TeV detection by HAWC.
  Given the source location and TS value (33), there may be a large
  contribution of the Galactic diffuse emission to this source.
  Multiwavelength catalog searches reveal several pulsars, several X-ray
  sources, and HII regions coincident with or in the vicinity of
  2HWC J1914+117*.
  There have been seven Swift observations, but the overall exposure is too low to
  identify a possible counterpart.  
  There are no possible counterparts in the \LAT catalogs. 

  The pulsars from the ATNF pulsar catalog located
  in the vicinity of 2HWC J1914+117* are:
  PSR J1915+1144 (0.1\degree, $d=7.2$\,kpc),
  PSR J1915+1149 (0.1\degree, $d=14$\,kpc),
  PSR J1913+1145 (0.2\degree, $\dot{E}=6.9\times 10^{33}$\ergps, $d=14$\,kpc, $\tau=967$\,kyr), and
  PSR B1911+11   (0.4\degree, $\dot{E}=1.2\times 10^{32}$\ergps, $d=3.1$\,kpc, $\tau=14.5$\,Myr).

\subsection{2HWC J1921+131}
  2HWC J1921+131 is a new TeV detection by HAWC.
  Given the source location and TS value (30.1), there may be a large
  contribution of the Galactic diffuse emission to this source.
  Multiwavelength catalog searches reveal several pulsars, several X-ray
  sources, and a molecular cloud system coincident with or in the vicinity of
  2HWC J1921+131. Swift observations exist of the source IGRJ19203+1328.
  There is no possible counterpart in the \LAT catalogs within a
  radius of 1\degree. 
  PSR J1919+1314 is the only nearby pulsar from the ATNF pulsar catalog,
  0.4\degree away. It is an old (2.4\,My) pulsar at a distance $d=13\,\textrm{kpc}$
  and not very energetic ($\dot{E}=8\times 10^{32}$\ergps), making the
  association unlikely.

\subsection{2HWC J1922+140 -- W51C}
  2HWC J1922+140 is associated with the radio-bright SNR W51C, which is
  located at a distance of $\sim$5.5\,kpc
  \citep{2010ApJ...720.1055S} and is a middle-aged remnant (${\sim} 3\times 10^4$\,yr) with an
  elliptical shape in radio encompassing a size of $0.6^\circ \times 0.8^\circ$
  \citep{1995ApJ...447..211K}.
  W51C was detected by \LAT in the energy range from 200\,MeV to 50\,GeV.
  \citet{2016ApJ...816..100J} reported a
  high-energy break in the energy spectrum of 2.7\,GeV and a spectral
  index beyond the break at $-2.52^{+0.07}_{-0.06}$.
  In \citet{2012A&A...541A..13A}, the MAGIC collaboration reported the detection
  of W51C at the $11\sigma$ level and a spectral index of
  $-2.58\pm0.07_{stat}\pm0.22_{sys}$. Above 1\,TeV, MAGIC observes W15C as an
  elongated region of half width about 0.1\degree on the long axis.
  
  2HWC J1922+140 is detected by HAWC in the point source search, however the 
  residual map exhibits various excess around the position of the source once
  the point source modeled has been subtracted.
  This indicated there may be additional emission farther away from W51C than
  previously reported.
  Given the source location, there may be a significant
  contribution of the Galactic diffuse emission to this extended emission.
  The spectrum fit is thus performed but using a point source model and an
  extended source model, with radius 0.9\degree.
  The spectrum measurement reported in Table~\ref{tab:fluxes} under the point
  source hypothesis appear to be in
  agreement with the MAGIC and \LAT results, while the one performed with the
  extended hypothesis is larger by about a factor 3.

  \subsection{2HWC J1928+177 and 2HWC J1930+188 region}
  
    In this region, two sources are found in the point search: 2HWC J1928+177
    and 2HWC J1930+188.
    Only the second source is previously detected in TeV, even though the
    location of the first source has been observed by IACTs.
    This region also exhibits signs of additional emission, which will be
    investigated in future analysis.
    
    2HWC J1928+177 is a new TeV source discovered in the point source search.  
    It is likely associated with the pulsar PSR J1928+1746
    (0.03\degree away, $\dot{E}=1.6\times 10^{36}$\ergps, $d=4.3$\,kpc, $\tau=83$\,kyr),
    the first pulsar discovered in the Arecibo L-band Feed Array (ALFA) survey
    \citep{2006ApJ...637..446C}.
    This pulsar and 2HWC J1928+177 are also within the 99\% uncertainty region of the
    unidentified EGRET source 3EG J1928+1746 which shows significant variability
    \citep{1999ApJS..123...79H}.
    The \LAT association for this EGRET source is 3FGL J1928.9+1739.
    However, the 3FGL source position and the 2HWC J1928+177 source position are
    not consistent within statistical uncertainty. 
    Also note that \LAT reported two analysis flags associated with this source,
    indicating a significant dependency of the reported source on the
    choice of the background model and other possible issues with detection or
    characterization of the source.
    VERITAS has also observed the location of PSR J1928+1746 \citep{2010ApJ...719L..69A}.
    However, VERITAS only observed a 1.2$\sigma$ excess at the source position, 
    and set a flux upper limit above 1\,TeV at the 99\% confidence level assuming
    a power law distribution with power law index of $-2.5$ at $2.6 \times 10^{-13}$\ifu.
    Even though the power law index assumed by VERITAS is similar to the HAWC measured
    spectral index, the flux measured by HAWC is about three times larger than the VERITAS limit,
    which seems to indicate that the spatial extent of PSR J1928+1746 is larger than the PSF of VERITAS.

    2HWC J1930+188 is associated with the supernova remnant SNR G054.1+00.3,
    which is a known TeV source discovered by VERITAS \citep{2010ApJ...719L..69A}.
    The VERITAS observation is consistent with a point-like source within
    the resolution of the instrument.
    SNR G054.1+00.3 hosts a young and energetic pulsar, PSR J1930+1852, at its center
    ($\dot{E}=1.2\times 10^{37}$\ergps, $d=7$\,kpc, $\tau=2.9$\,kyr).
    \citet{ExicXRay} reported the discovery of a nonthermal  X-ray jet that is consistent with a radio extension.
    It confirms the existence of a PWN in the SNR G054.1+00.3.
    The spectral indices and fluxes at 7\,TeV of VERITAS and HAWC are consistent
    within statistical and systematic uncertainties.
    The HAWC measurements indicate that the TeV spectrum associated with
    SNR G054.1+00.3 extends beyond the VERITAS measured energy range (250\,GeV -- 4\,TeV).
    
    As explained in Section \ref{sec:results-catalog}, the flux has also been
    calculated under an extended source hypothesis. 
    The radius has been chosen to include the region around 2HWC J1928+177 and
    2HWC J1930+188.
    Table~\ref{tab:fluxes} shows that the measured flux for this whole
    region is significantly larger than the sum of the fluxes of 2HWC J1928+177
    and 2HWC J1930+188 under the point source hypothesis, thus favoring
    extended emission or additional unresolved sources.
    
   \subsection{2HWC J1938+238}
   2HWC J1938+238 is a new TeV source discovered in the point source search,
   within the Galactic Plane.
   There are several optical galaxies, radio galaxies, and an ATNF pulsar within
   0.5\degree around the source location.
   However, none of these sources are known X-ray or gamma-ray sources.
   The pulsar, PSR J1940+2337, is located 0.4\degree away from 2HWC J1938+238
   and is a middle age pulsar (113
   \,kyr) with a spindown power $\dot{E}=1.9\times 10^{34}$\ergps and a
   distance $d=8.5$\,kpc.
   
   \subsection{2HWC J1949+244}
   2HWC J1949+244 is a new TeV source discovered within the Galactic Plane.
   The source is discovered in the 1\degree extended search, which, given the
   low latitude of the source, suggests there can be an important contribution
   of the Galactic diffuse emission to this source.
   It is located 0.1\degree away from the 
   unidentified \LAT source 3FGL J1949.3+2433.
   The extent of 3FGL J1949.3+2433 is less than 0.1\degree, which is much smaller
   than the size of the search in which 2HWC J1949+244 was found.
   The \LAT measured spectral index of this source is $-2.8 \pm 0.2$, which is
   slightly softer than the one measured by HAWC.
   
   The millisecond pulsar PSR J1950+2414 is also located near 2HWC J1949+244
   (0.3\degree, $\dot{E}=9.4\times 10^{33}$\ergps, $d=7.3$\,kpc, $\tau=3.6$\,Gyr).
   However, this source has not been detected in X-ray or GeV \citep{2015ApJ...806..140K}.
   
   \subsection{2HWC J1953+294 and 2HWC J1955+285 region} 
   In this region, two sources are found nearby in the point source search:
   2HWC J1953+294 and 2HWC J1955+285, none of which has previous TeV detection.
   
   After the HAWC discovery of 2HWC J1953+294, VERITAS observed this source for 37
   hours and confirmed the existence of the TeV source.
   The VERITAS observations of this source will be continued during the 2016--2017 
   season \citep{Holder:2016heb}.
   2HWC J1953+294 is located at 0.2\degree from the pulsar wind nebula DA 495, which is
   associated with the supernova remnant G65.7+1.2.
   It is likely that the 3FGL J1951.6+2926 is associated with the central pulsar
   of this system \citep{Karpovaetal}.
   A joint analysis of this region with \LAT, VERITAS, and HAWC data is ongoing.
   
   The second new source, 2HWC J1955+285, may be associated with the shell-type
   supernova remnant SNR G065.1+00.6, located 0.5\degree away.
 The first gamma-ray source in the region of SNR G065.1+00.6 was reported by the COS-B
 satellite as 2CG 065+00 \citep{COSBCat2}, then confirmed by the EGRET detection 3EG J1958+2909 \citep{3EGCatalog}.
  2HWC J1955+285 is near the energetic \LAT pulsar PSR J1954+2836
  (0.2\degree away, $\dot{E}=1.0 \times 10^{36}$\ergps, $\tau=69$\,kyr).
  \LAT also reported a nonobservation of the SNR in \citet{2016ApJS..224....8A}.
  Milagro reported a 4.3$\sigma$ excess at this location \citep{Abdo:2009ku}.
  MAGIC reported a non detection and set a flux limit at 2--3\% of the Crab Nebula flux
  at 1\,TeV \citep{Aleksic:2010dh}.

\subsection{Cygnus region}
Within Galactic longitude 70\degree and 85\degree in the Galactic
plane, there are five 2HWC sources.  One is potentially part of the
very extended emission in the Cygnus Cocoon field, and the rest are mostly
associated with known TeV gamma-ray sources.

2HWC J2006+341 is observed with a TS value of 36.9 and is
unassociated with any known TeV detections.
Milagro has reported a 3.3$\sigma$ excess at this location.
The nearest gamma-ray source is 0.7\degree
away, an unidentified \LAT source 3FGL J2004.4+3338.
This source was also reported in the 1FHL catalog but not the
2FHL catalog.  Within a 1\degree radius there are no nearby SNRs from
the Manitoba catalog.
The nearest pulsar from the ATNF pulsar catalog is PSR J2004+3429, 0.4\degree away.
Its characteristics are $d=11$\,kpc, $\dot{E}=5.8\times 10^{35}$\ergps, and a
characteristic age of 18\,kyr.

2HWC J2019+367 is associated with MGRO J2019+37,
which has a reported extent of 0.7\degree from a 2D Gaussian fit
\citep{MGRO-cygnusspec}.
The extended Milagro source is resolved into two by VERITAS
\citep{VERJ2019p367}, VER J2016+371 and VER J2019+368, with brighter
emission coming from the latter.
The nature of VER J2016+371 is unclear and could be associated with
either the supernova remnant CTB~87 or a blazar, both have been
detected by \LAT.
VER J2019+368 is extended and encompasses two pulsars, PSR J2021+3651
(3FGL J2021.1+3651) and PSR J2017+3625 (3FGL J2017.9+3627), and a star
forming region Sh~2-104 that could all contribute to the extended TeV
emission \citep{NuSTARsh2104}.  The spectrum of VER J2019+368 is
derived from a circular region of 0.5\degree radius and is very hard
with a photon index of $-1.75\pm 0.3$ up to 30\,TeV.  Comparing the
integrated flux between 1 and 30\,TeV, the 2HWC measurement from a
point source assumption is still higher than that of the VERITAS extended
assumption.  The PSF of this HAWC dataset below 1\,TeV
is more extended than the 0.5\degree radius used by VERITAS and
the source could be more extended than previously thought.
The integrated flux from the extended source fit of the
HAWC source is more consistent with the Milagro measurement.

2HWC J2020+403 is likely associated with
VER J2019+407 \citep{VER-G78p2}.  TeV emission from this source is
unidentified and is potentially associated with the supernova remnant
G78.2+2.1 \citep[e.g.][]{2016ApJ...826...31F} or the gamma-ray pulsar
PSR J2021+4026 ($\dot{E}=1.2\times 10^{35}$\ergps, $d=2.1$\,kpc, $\tau=77$\,kyr).
The supernova remnant G78.2+2.1 (Gamma Cygni) is
detected as extended by \LAT and reported in both the 3FGL and the
2FHL catalogs.
The flux observed by HAWC is higher
than the one reported from VER J2019+407.  HAWC may be
measuring multiple emission components.

Diffuse emission in this region with a 2D Gaussian width of $(2.0 \pm
0.2)$\degree has been reported by the Fermi collaboration
\citep{Fermi-cocoon}.  The GeV diffuse emission is named the Cygnus
Cocoon, and likely originates from a superbubble of freshly
accelerated cosmic rays that are confined up to 150\,TeV. 
ARGO J2031+4157 is reported as the counterpart of
the Cygnus Cocoon \citep{ARGOcocoon}.  The 2D Gaussian width of this
source is measured to be $(1.8 \pm 0.5)$\degree after subtraction of
nearby known TeV sources.  This is in agreement with the extended emission
reported by Milagro, which has a 2D Gaussian width of
1.8\degree and a spectrum compatible with
an extrapolation of the Fermi Cocoon spectrum \citep{MGRO-cygnusspec}.

2HWC J2031+415 is associated with TeV J2031+4130, a PWN
first reported as unidentified in TeV by HEGRA \citep{J2032HEGRA}.
Various IACTs have reported
pointlike or up to 0.2\degree extended emissions from the pulsar
position with consistent
spectra \citep{J2032Whipple,J2032MAGIC,J2032VERITAS}, while Milagro
and ARGO have reported extended emission compatible with the Cygnus
Cocoon as mentioned above.
The HAWC flux is more consistent with the flux measured by Milagro and
ARGO than the IACTs, in agreement with possible additional emission components
besides the PWN within the region.

2HWC J2024+417* is detected with $\mathrm{TS}=28.4$ and could be part
of the extended morphology of 2HWC J2031+415.  It is 0.35\degree from
3FGL J2023.5+4126, which is associated with the Cygnus Cocoon field.
In addition to the diffuse emission, the 3FGL catalog also lists
multiple sources associated with the Cygnus Cocoon field.

\section{Source population} 
  \label{sec:population}
  A total of 39 sources are identified in the catalog.
Two are associated as blazars, two as SNRs, seven as PWNe, and 14 other have possible
associations with PWN, SNR, and molecular clouds.
The remaining 14 are unassociated.

The majority of the sources in the catalog lie near the Galactic Plane.
Figure \ref{fig:galacticlb} illustrates the distributions of the sources in Galactic latitude $b$ and longitude $l$,
as well as the sensitivity, for sources within 10$^{\circ}$ from the Galactic Plane.
It can be seen that our sensitivity is highly uniform in a wide band around the
Galactic Plane ($-10^{\circ} < b < 10^{\circ}$), which is in contrast to IACT
surveys \citep[see e.g.][]{2006ApJ...636..777A}.
The close to uniform sensitivity in $b$ of this
catalog ensures completeness even for (likely nearby) Galactic sources at
moderate to large galactic latitudes.  Indeed two new sources are found at
rather large galactic latitudes: 2HWC J0700+143 at $b=8.44^{\circ}$
and 2HWC J1829+070 at $b=8.09^{\circ}$). However, the distribution
of the newly observed sources peaks within $|b| < 1^{\circ}$. In
Figure \ref{fig:galacticlb}, the total and new 2HWC source distributions are
compared to the known distributions of supernova remnants from
\citet{2014BASI...42...47G} and pulsars with a spindown
luminosity $\dot{E} > 10^{34}$\ergps from
\citet{2005AJ....129.1993M}. When taking into account the sensitivity
of this catalog in $l$, the distribution of the new sources is broadly 
consistent with that of known SNRs and PSRs.

 \begin{figure*}
    \centering
    \includegraphics[width=0.338\textwidth]{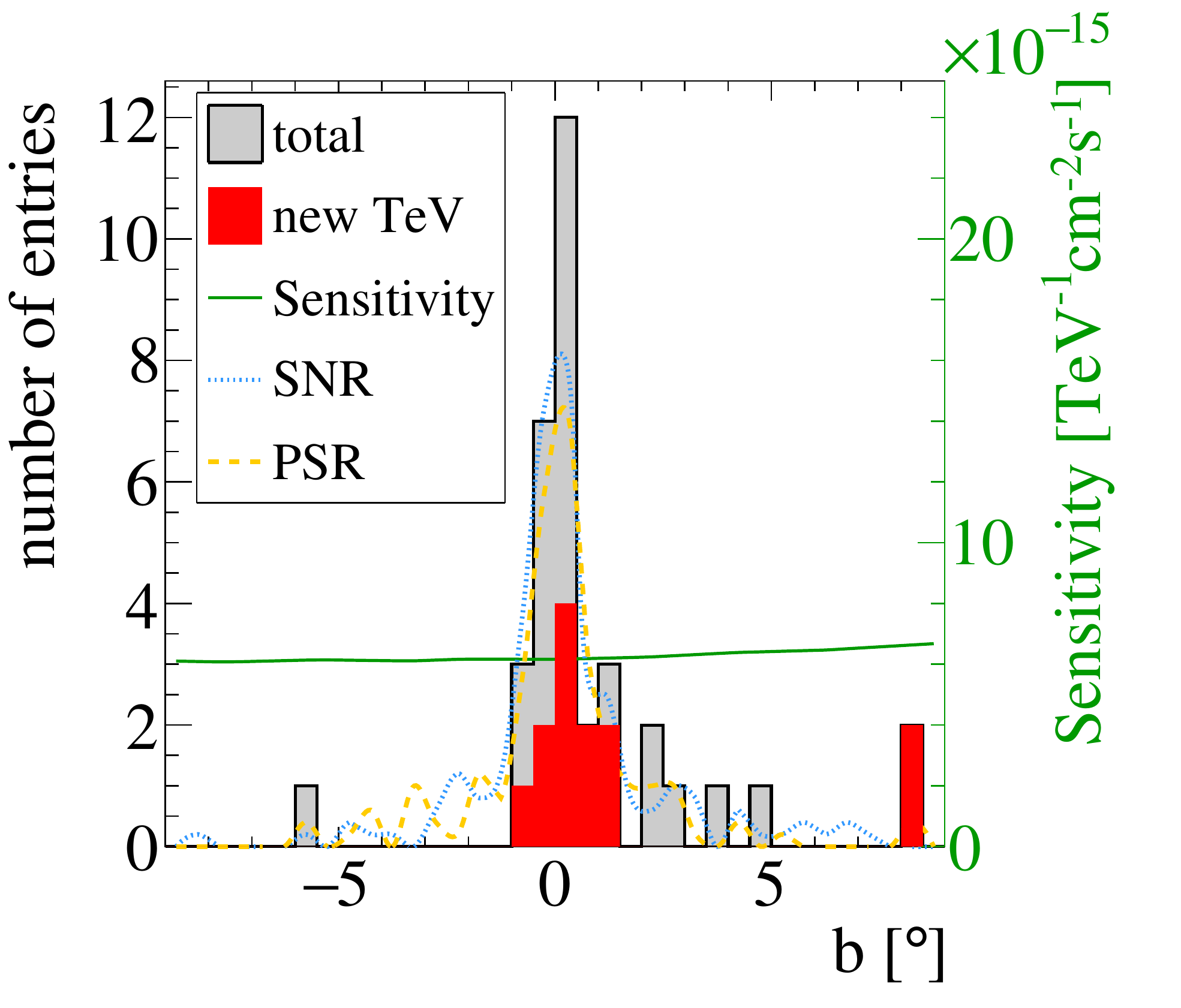}
    \includegraphics[width=0.605\textwidth]{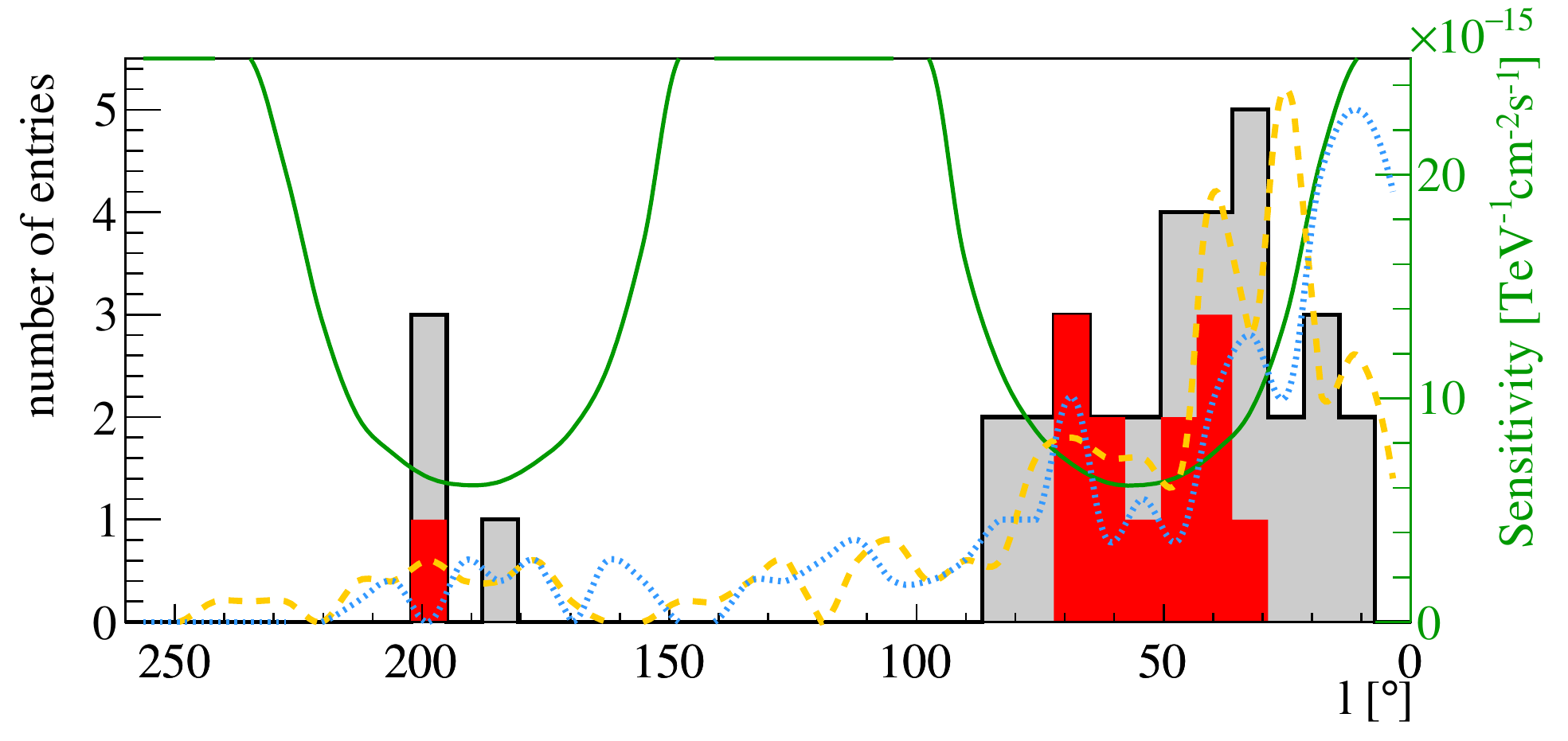}
    \caption{\emph{Left}: Galactic latitude distribution of 2HWC catalog sources
      in bins of $\Delta b = 0.5^{\circ}$.  \emph{Right}:
      Galactic longitude distribution in bins of
      $\Delta l = 7.2^{\circ}$. The subset of sources without a 
TeVCat association are shown in red. The right-hand axis on
the plot indicate the differential point-source flux sensitivity of the survey
at 7~TeV.
     In the case
      of the $b$-distribution, the sensitivity at $l=60^{\circ}$ is
      indicated by the green line and for the $l$-distribution the
      sensitivity is shown for $b=0^\circ$. Both distributions are
      compared to distributions of known pulsars
      \citep{2005AJ....129.1993M} and supernova remnants
      \citep{2014BASI...42...47G} in the field of view of HAWC. Both
      pulsars and supernova remnants distributions
      are binned in the same way as the 2HWC sources and re-scaled for ease of
      comparison. In addition, only pulsars with a
      spindown luminosity of $\dot{E} > 10^{34}$\ergps are indicated.}
 \label{fig:galacticlb}
  \end{figure*}
  
As noted earlier, in the Inner Galactic Plane, the Galactic diffuse emission may have a
significant impact the flux measurement of some sources near the TS threshold.
The current knowlege of this emission in the TeV regime is limited, and HAWC
is uniquely suited to measure this Galactic diffuse emission in the future.

Out of the Galactic Plane, 2HWC J1104+381 (Mrk421) and 2HWC J1653+397 (Mrk 501)
are the only sources with known extragalactic association.
We also identify four sources, which have no association, but are also very
close to the TS threshold indicating that they may be statistical fluctuations.
Random fluctuations are expected to appear mostly out of the Galactic Plane
since the latter only represents a small fraction of the sky.
However, the expected number of false positive in the catalog search is 0.5, so
we regard these sources as interesting and certainly worthy of further scrutiny.
Overall, the \extragal sources represent a smaller fraction of the total number
of sources than typically observed by other gamma-ray instruments
(e.g. $>$75\% of extragalactic sources in \LAT 2FHL, and about 50\% for IACTs
in TeVCat).
This is due to the sensitivity of HAWC peaking at higher energy than satellites
and IACTs, energy where VHE gamma rays are attenuated by interaction with
the extragalactic background light (EBL).

\section{Conclusions}

  \label{sec:outlook}
  The 2HWC catalog is the result of the first search performed with 507 days of
data from the fully deployed HAWC Observatory.
It is the most sensitive unbiased TeV survey of large regions of the northern
sky performed to date.
The peak sensitivity of this survey lies around 10\,TeV, depending on the
source spectrum.
This allowed the detection of a total of 39 sources, 16 of which are more than a degree
away from sources reported in TeVCat.
The source characteristics (location, spectrum, and for some a tentative
indication of the extent) were presented, and possible associations were
discussed.
Twenty-eight sources have no firm associations.
Some are in complex regions with nearby sources and refined analysis as well
as more statistics will help the source identification.
Four sources are found in the extended search only.

HAWC is continuously taking data and the analysis and detector modeling are
being refined.
Future analyses will include more data, explore the modeling of multiple
sources and of detailed morphologies making use of multi-instrument and
multiwavelength information.

  \vspace{1em}
  We acknowledge the support from: the US National Science Foundation (NSF);
the US Department of Energy Office of High-Energy Physics; the Laboratory
Directed Research and Development (LDRD) program of Los Alamos National
Laboratory; Consejo Nacional de Ciencia y Tecnolog\'{\i}a (CONACyT), M{\'e}xico
(grants 271051, 232656, 260378, 179588, 239762, 254964, 271737, 258865, 243290,
132197), Laboratorio Nacional HAWC de rayos gamma; L'OREAL Fellowship for Women
in Science 2014; Red HAWC, M{\'e}xico; DGAPA-UNAM (grants RG100414, IN111315,
IN111716-3, IA102715, 109916, IA102917); VIEP-BUAP; PIFI 2012, 2013, PROFOCIE
2014, 2015; the University of Wisconsin Alumni Research Foundation; the
Institute of Geophysics, Planetary Physics, and Signatures at Los Alamos
National Laboratory; Polish Science Centre grant DEC-2014/13/B/ST9/945;
Coordinaci{\'o}n de la Investigaci{\'o}n Cient\'{\i}fica de la Universidad
Michoacana. Thanks to Luciano D\'{\i}az and Eduardo Murrieta for technical
support.

\bibliography{catalogref}

\end{document}